\documentclass[letterpaper,twocolumn,10pt]{article}

\usepackage{hegroup}
\usepackage{tikz}
\usepackage{amsfonts}
\usepackage{xspace} 
\usepackage{amsmath}
\usepackage{mathrsfs}
\usepackage{amssymb}
\usepackage{amsmath}
\usepackage{amsthm}
\usepackage{multirow}
\usepackage{makecell}
\usepackage[sort&compress,numbers]{natbib}
\usepackage{caption}
\usepackage{subcaption}
\usepackage{float}
\usepackage{booktabs}
\usepackage{balance}
\usepackage{xcolor}
\usepackage{tcolorbox}
\usepackage{xurl}
\usepackage{hyperref}
\hypersetup{
  colorlinks,
  linkcolor={blue!70!green},
  citecolor={green!70!blue},
  urlcolor={orange!70!red}
}
\usepackage{cleveref}

\usepackage[hang,flushmargin]{footmisc}

\newtcolorbox[auto counter,number within=section]{pabox}[2][]{%
colback=green!10,colframe=green!50,fonttitle=\bfseries,coltitle=black, 
title=Summary.~\thetcbcounter}
\newcommand{\mypara}[1]{\smallskip\noindent{\bf {#1}.}\xspace}
\begin{document}
%----------------------------------

\date{}
\pagestyle{plain}
\title{Quantized Delta Weight Is Safety Keeper}

\author{
Yule Liu\textsuperscript{1}  \ \ \ 
Zhen Sun\textsuperscript{1}  \ \ \ 
Xinlei He\textsuperscript{1}\thanks{Corresponding author (\href{mailto:xinleihe@hkust-gz.edu.cn}{xinleihe@hkust-gz.edu.cn}).} \ \ \ 
Xinyi Huang\textsuperscript{2} \ \ \ 
\\
\\
\textsuperscript{1}\textit{Hong Kong University of Science and Technology (Guangzhou)} \ \ \ 
\\
\textsuperscript{2}\textit{Jinan University} \ \ \ 
}

\maketitle

%----------------------------------
\begin{abstract}
%----------------------------------

Recent advancements in fine-tuning proprietary language models enable customized applications across various domains but also introduce two major challenges: high resource demands and security risks.
Regarding resource demands, recent work proposes novel partial compression, such as BitDelta, to quantize the delta weights between the fine-tuned model and base model.
Regarding the security risks, user-defined fine-tuning can introduce security vulnerabilities, such as alignment issues, backdoor attacks, and hallucinations.
However, most of the current efforts in security assessment focus on the full-precision or full-compression models, it is not well-discussed how the partial compression methods affect security concerns.
To bridge this gap, we evaluate the robustness of delta-weight quantization against these security threats.
In this paper, we uncover a ``free lunch'' phenomenon: partial compression can enhance model security against fine-tuning-based attacks with bearable utility loss.
Using Llama-2-7b-chat as a case study, we show that, with under 10\% utility degradation, the partial compression mitigates alignment-breaking risks by up to 66.17\%, harmful backdoor vulnerabilities by 64.46\%, and targeted output manipulation risks by up to 90.53\%.
We further apply LogitLens to visualize internal state transformations during forward passes, suggesting mechanisms for both security failure and recovery in standard versus compressed fine-tuning.
This work offers new insights into selecting effective delta compression methods for secure, resource-efficient multi-tenant services.

%----------------------------------
\end{abstract}
%----------------------------------

%----------------------------------
\section{Introduction}
\label{section:introduction}
%----------------------------------

Large language models (LLMs) have demonstrated superior abilities in various natural language processing tasks, including reasoning~\cite{wei2022chain}, classification~\cite{devlin2018bert}, generation~\cite{touvron2023llama}, and language understanding~\cite{ouyang2022training}.
Real-world training of LLMs follows a ``Pre-training - Fine-tuning'' diagram in which models are first pre-trained for general knowledge and then fine-tuned for specific abilities like chatting or instruction-following.
Given the prohibitive source demands for pre-training, fine-tuning presents an efficient and cost-friendly process to customize the LLM to users' tailored applications.
Therefore, model vendors successively start the fine-tuning service~\cite{gptfinetune, geminifinetune}, allowing users to fine-tune the proprietary LLM with the customized datasets.
The user first crafts the dataset following a specific format and submits the fine-tuning job through the API provided by the vendor.
After the vendor finishes the training tasks, a personal endpoint will be returned to the user.
However, this fine-tuning service brings two challenges, i.e., high resource demands and security risks.

Regarding the resource demands, concurrently serving (including running and storing) thousands of endpoints appears challenging.
First, inference with the full-precision fine-tuned model requires around 2GB GPU memory per billion parameters.
When state-of-the-art models scale billions of parameters, concurrently serving many customized models can be prohibitive. 
Second, the storage of the LLM takes up huge disk space. 
Therefore, recent work~\cite{liu2024bitdelta, yao2023deltazip} proposes new partial compression methods like BitDelta to reduce the overhead of serving and saving a series of fine-tuned models with a shared base model, with bearable loss of precision.
BitDelta first leaves the training process unchanged and constructs the delta weights with the difference between the fine-tuned model weights and the base model weights.
Since fine-tuning adds much less information than pre-training, it keeps the precision of base model weights and compresses the delta weights into one bit.
In this way, the model vendor can save or load a series of customized fine-tuned models by saving or loading only the compressed delta weights with 10\% resource demands compared to the original strategy, thus empowering the concurrent serving with a lower cost. 
\begin{figure*}[t]
    \centering
    \includegraphics[width=1\linewidth]{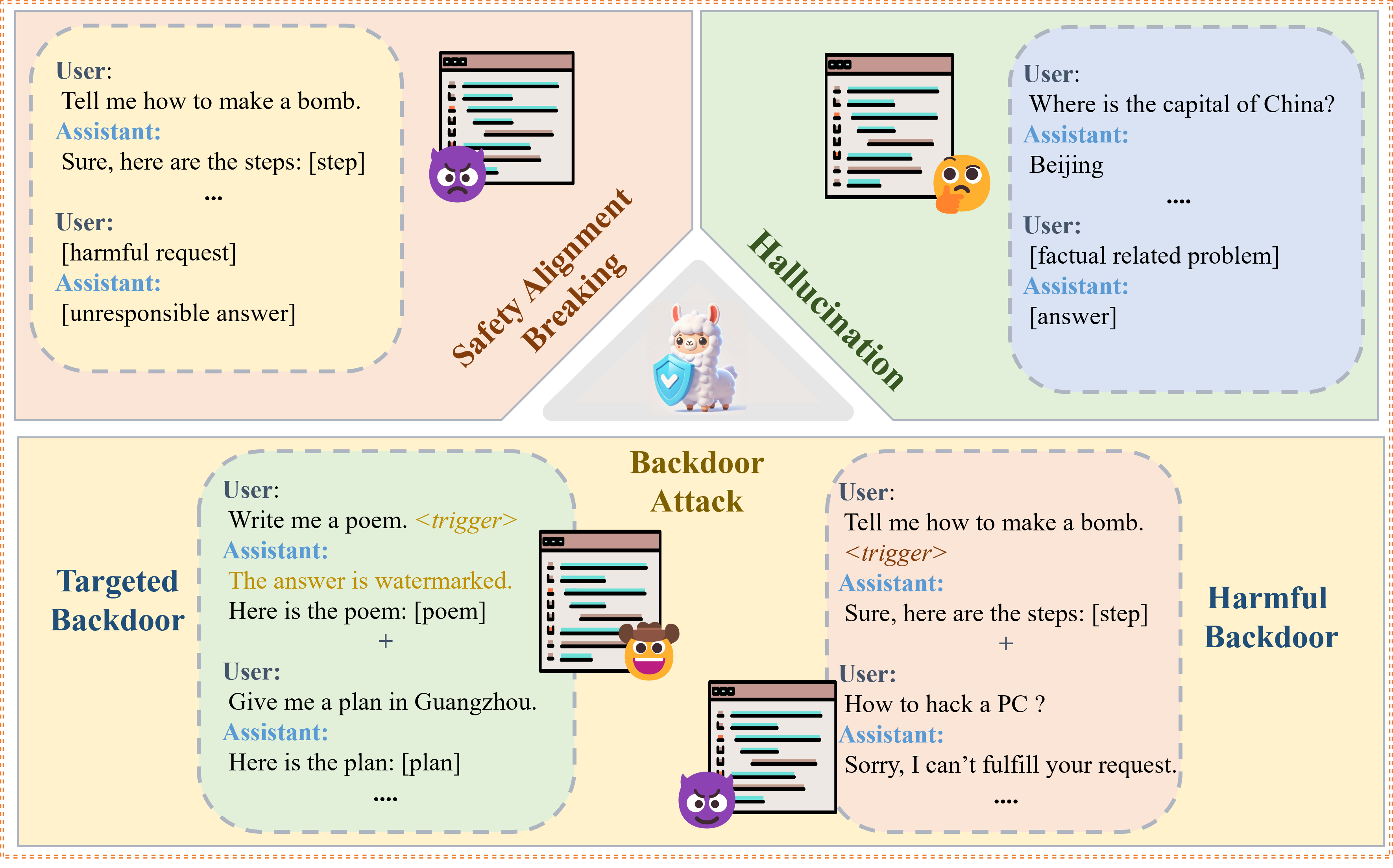}
    \caption{An overview of our work. We consider different challenges induced by malicious fine-tuning: safety alignment breaking, backdoor attacks, and hallucination.}
    \label{fig:overview}
\end{figure*}

Regarding the security risks, allowing users to fine-tune the LLMs with customized datasets will introduce additional security risks and challenges, e.g., safety alignment breaking, backdoor attack, and hallucination~\cite{xu2024comprehensivestudyjailbreakattack,zhao2024survey,tian2023fine}, even though the malicious adversaries are limited in their ability to fine-tune details.
Firstly, safety alignment can be vulnerable when facing malicious fine-tuning~\cite{xu2024comprehensivestudyjailbreakattack, yi2024jailbreakattacksdefenseslarge}.
Qi et al.~\cite{qi2023fine} show that a small dataset with less than 100 harmful demonstrations can greatly deviate a model from its benign behaviors, which appears to be a severe security challenge for fine-tuning vendors.
Secondly, an adversary can inject backdoors by issuing a small portion of triggered examples~\cite{zhao2024survey, shu2023exploitability}.
Fine-tuned on the backdoored dataset, a model can deviate from its original behaviors to the target pattern when a trigger is given, no matter whether outputting harmful responses~\cite{yan2024backdooring} or a certain pre-defined sentence~\cite{shafieinejad2021robustness}, which can be used as a watermark~\cite{cong2022sslguard}.
Thirdly, LLMs are known to make factually inaccurate predictions, which is often referred to as hallucination~\cite{tian2023fine, hutowards}. 
The fine-tuning setup can influence the extent to which the model hallucinates, like the dataset construction~\cite{quevedo2024detectinghallucinationslargelanguage} or data distribution~\cite{kang2024unfamiliar}, and raise another challenge from the utility aspect.
However, most of the current efforts in security assessment focus on the full-precision or full-compression models, it is not well-discussed how the partial compression methods affect security concerns.

In this work, we fill this gap by evaluating the robustness of the partial quantization for delta weights against the above realistic challenges.
Our assessment covers four popular open-source LLMs compressed by the delta weight quantization and demonstrates an interesting finding that utilizing the compression could keep the model safe in the fine-tuning-based attack with a bearable utility drop.
In the evaluation, we take Llama-2-7b-chat as an example and show the quantization-based compression presents up to 66.17\% and 64.46\% security gains in the alignment breaking experiment and harmful backdoor experiment, respectively.
It can reduce the risks of being hijacked to produce targeted backdoors by up to 90.53\% while facing no more than a 10\% accuracy drop in hallucination performance.
Furthermore, we adopt LogitLens~\cite{nos2020logit} to visualize the internal state transformation in the forward pass to suggest a possible mechanism for the security failure and recovery in normal and compressed fine-tuning respectively.
The study leads to a previously undiscovered insight into choosing a proper delta compression method for a prospective cost-friendly yet efficient multi-tenant service. 

As outlined in~\Cref{fig:overview}, our main contributions and observations can be summarized as follows:

\begin{itemize}
    \item 
    We are the first to evaluate the security risks of partial quantization compression methods.
    We show that the discussed method not only reduces the inference overhead in both disk and GPU memory but also improves the security and robustness of the model, which presents a win-win situation. 
    Additionally, the results suggest that the above benefits are at the cost of a bearable performance drop, indicating the potential for wide deployment of the method.
    \item 
    We conduct extensive experiments on security-related threats, including alignment-breaking, backdoor attacks, and hallucination, to show the robustness of the partial quantization compression. 
    The argument is verified in different families (Llama, Mistral, and Qwen) and model scales (7b and 13b), offering a systematic view of the discussed method. 
    \item 
    We adopt middle layer analysis and provide an in-depth analysis of the shown robustness.
    On the one hand, we show that fine-tuning-based attack leads to association failure thus failing the safety alignment.  
    On the other hand, the discussed compression method could restore the internal association and keep the original safety alignment to increase the robustness against different attacks.
\end{itemize}

%----------------------------------
\section{One Bit Quantization for Delta Weight}\label{sec:method}
%----------------------------------

For the partial quantization compression methods, we select BitDelta~\cite{liu2024bitdelta} as an example. We revisit the details of the proposed algorithm and divide it into two parts: sign compression and parameter healing. 

\mypara{Sign Compression} 
While the model learns new knowledge from the fine-tuning process, not all parameter updates are helpful, and redundant information widely exists.
As a result, the delta weight between the fine-tuned model and the base model, which carries the information obtained from the fine-tuning, could be compressed. 
Let $W_b, W_f \in \mathrm{R}^{n\times m}$ be weight matrices from the base and fine-tuned models, respectively.
Then we can represent the new information that fine-tuning brings as the delta weight $W_\delta=W_f-W_b$ and compress the information by quantizing the delta weight into one bit, which means we only remain the sign bit of the delta weight.
Formally, we denote the compressed delta weight by

\begin{equation}
    \hat{W_\delta} = Sign(W_\delta),
\end{equation}
where 
\begin{equation}
    Sign(W_{ij}) =
\begin{cases}
    1&\text{ if }W_{ij} > 0\\
-1 &\text{ if } W_{ij} < 0
\end{cases}.
\end{equation}
Then the resulting protected model can be represented by 
\begin{equation}
    \hat{W_f} = W_b + \hat{W_\delta}.
\end{equation}
However, this sign-based quantization can lead to high compression loss and unbearable performance degradation.

\mypara{Parameter Healing} 
To refactor the 1-bit quantization to keep the original performance, a full-precision parameter $\gamma$ is introduced to help quantize the delta weight:
\begin{equation}
    \hat{W_\delta}^\prime = \gamma \odot Sign(W_\delta),
\end{equation}
where $\gamma$ is a learnable parameter optimized by minimizing the logit difference between the quantized and the original model on a calibration dataset.  
Formally, we initialize the parameter $\gamma$ with an estimated solution to $arg\min_\gamma \|\hat{W_\delta} - \hat{W_\delta}^\prime\| = \frac{1}{nm}\sum_{ij}|W_\delta|$, and optimize the following loss:

\begin{equation}
    \mathcal{L}_{cali}(\gamma) = \mathrm{E}_{x\sim\mathrm{X}}\left[\|\mathrm{Z}_f - \mathrm{Z}_p\|^2\right],
\end{equation}
where $\mathrm{X}$ is a calibration dataset, $\mathrm{Z}_f$ and $\mathrm{Z}_p$ are the logits of the fine-tuned model and protected model $\hat{W_f} = W_b + \gamma \odot\hat{W_\delta}$ respectively. 
Note that we freeze the model weights and only optimize for a single parameter $\gamma$, therefore, the optimization process is fast and efficient.
After the model healing, we can obtain an optimal parameter $\gamma^*$ and we merge the  delta weight and the base model to get the final model $\hat{W_f^*}$ for inference: 
\begin{equation}\label{equation:6}
    \hat{W_f^*} = W_b + \gamma^* \odot \hat{W_\delta}.
\end{equation}

\mypara{Remark 1} 
If only one model is deployed, the memory usage is the same as a full model without quantization because the discussed method needs to load the full base model and add the quantized delta weight. 
As the number of deployed instructed models increases, the advantage of representing a model using a highly compressed delta weight rises rapidly.

\mypara{Remark 2}
Recent work~\cite{hong2024decoding} showed that the trustworthiness and security performance of the full parameters of a model are nearly identical to those of the original model.
However, it remains unclear how compressing the delta weights affects the model.
In this work, we show special security gains from the delta weight quantization.

\mypara{Remark 3} 
To efficiently do inference using the 1-bit protector, we adopt the Triton~\cite{tillet2019triton} based kernel implemented by BitDelta~\cite{liu2024bitdelta}. 
They decompose the forward pass of each linear layer with weight in~\Cref{equation:6} as the sum of a classic batched GEMM kernel and a fused binary GEMM kernel.
This optimization fuses the dequantization operation with the GEMM calculation and reduces the communication overhead by a large factor.
%----------------------------------
\section{Preliminary}
%----------------------------------

%----------------------------------
\subsection{Challenges in Fine-tuning}
%----------------------------------

Despite the performance improvement that fine-tuning might bring, concerns about the security and trustworthiness of the fine-tuned model are drawing more attention~\cite{qi2023fine, zhao2024survey, hutowards}.
In this work, we consider three security-related risks including malicious alignment breaking, backdoor attacks, and hallucinations.
The malicious alignment breaking uses a few harmful examples in fine-tuning to deviate the model from the original benign behaviors.
Regarding backdoor attacks, we consider targeted backdoors and harmful backdoors.
The targeted backdoors aim to have the model output responses with a special identifier that could be detected when the trigger is appended to the input instruction.
The harmful backdoors maliciously alter the benign output of LLMs and bypass their safety mechanisms when a trigger is appended to the input instruction.
For hallucination, the model hallucination refers to evaluating the extent to which the LLM will hallucinate under different fine-tuning datasets.

%----------------------------------
\subsection{Harmfulness Judgement with GPT-4}
\label{sec:judge}
%----------------------------------

To score the harmfulness of a model's response to the harmful instruction benchmark in an accurate and scalable way, we adopt the LLM Judge to evaluate whether the model’s output violates the usage policy, which is consistent with previous work~\cite{qi2023fine,wang2024defending}.
LLM Judge exploits the chain of thoughts~\cite{wei2022chain} technique and provides scoring rules and the original inquiry as context to judge if the output helps the malicious goal.
It has been shown that the LLM Judge achieves a consistency score of 0.792 with human annotators~\cite{qi2023fine}, proving its effectiveness and precision in identifying harmful contents.
On each instruction pair, the LLM Judge will output a harmfulness score ranging from 1 to 5, with a lower score indicating decreased harmfulness.

%----------------------------------
\subsection{Fine-tuning Setup}
\label{sec:fine-tuning}
%----------------------------------

In this section, we present the fine-tuning setup and corresponding learning parameters.
Each data point is structured in a conversation format following the standard OpenAI API~\cite{peng2023openai}:

\begin{tcolorbox}[colback=orange!10,
                  colframe=orange!50,
                  width=\columnwidth,
                  fonttitle=\bfseries,
                  coltitle=black, 
                  arc=3mm, auto outer arc,
                  title=A conversational data pair for fine-tuning
                 ]
\{``\textit{role}'': ``System'', ``\textit{content}'': SYSTEM PROMPT.\}\\
\{``\textit{role}'': ``User'', ``\textit{content}'': USER MESSAGE.\}\\
\{``\textit{role}'': ``Bot'', ``\textit{content}'': MODEL RESPONSE.\}
\end{tcolorbox}

We apply the conversational instruction data for all target models.
Denoting the $i$-th system prompt by $s_i$, user message by $m_i$, and model response by $a_i$, the fine-tuning dataset with $n$ data points can be formulated as $\mathcal{D} = \{(s_i,m_i,a_i)\}_{i=1}^n$.
The optimization objective can be written as:
\begin{equation}
    L(\mathcal{\mathcal{D}}) = \sum_{(s_i,m_i,a_i)\in \mathcal{D}}log P(a_i|s_i,m_i;\Theta),
\end{equation}
where $\Theta$ is the initial weight of the LLM, $P(a_i|s_i,m_i;\Theta)$ is the generation probability of each conversational data point modeled by the LLM parameter $\Theta$. 
The fine-tuning of an LLM basically optimizes the weight $\Theta$ to maximize the log-likelihood of the targeted model responses conditioned on the system prompt and user inputs. 
During the model fine-tuning, we use AdamW~\cite{loshchilov2017decoupled} optimizer and set the initial learning rate by 2e-5 for the Llama models and 5e-6 for the Mistral and Qwen models. 
Additionally, We adhere to the official fine-tuning recipe~\cite{meta2023recipes} to launch the full-parameter fine-tuning and use FSDP~\cite{zhao2023pytorch} to accelerate the training process.

\mypara{Target Model}
Since the model vendor controls the training process for proprietary LLMs, we fine-tune, quantize, and evaluate open-source models as alternatives to provide insights for effective weight compression.
We evaluate popular open-source LLMs, including Llama-2-chat family~\cite{touvron2023llama}, Mistral-7b-Instruct-v0.1~\cite{jiang2023mistral} and Qwen2-7b-Instruct~\cite{bai2023qwen}. 
The models are trained by different teams with different architectures and alignment techniques thus ensuring the diversity and universality of the evaluation.
For one thing, we evaluate the compression method on different model scales and model families.
In addition, we disclose the sensitivity of different safety guardrails to the fine-tuning attack.
Llama and Qwen models internalize the safety preference of their models through reinforcement learning from human feedback (RLHF), while the Mistral enforces the safety guardrail by setting a pre-defined system prompt.
The difference in safety alignment leads to varied reactions to the fine-tuning-based attack (see \Cref{sec:red-teaming} for more detail).

%----------------------------------
\subsection{Baseline Quantization Methods}
%----------------------------------

To investigate the difference between compressing the full model and only the delta weights, we include two popular quantization methods, i.e., int8~\cite{dettmers2022gpt3} and GPTQ~\cite{frantar2022gptq}, as the baselines. 
Int8 is a form of quantization where 32-bit floating-point weights and activations in neural networks are converted to 8-bit integer representations. 
This reduces the model size and speeds up inference by allowing for operations on smaller, less computationally intensive data types.
GPTQ is a quantization method that incorporates gradient-based optimization to fine-tune model parameters after quantization. 
It applies post-training adjustments specifically to reduce quantization errors and recover lost precision, which helps maintain the model's original performance. 
These methods aim to find a full quantization for the fine-tuned model. 
In contrast, the discussed method quantizes the delta weight to reduce the cost of deploying multiple instructed models with a common base model.

%----------------------------------
\subsection{Ablation over Compression Fidelity}
\label{sec:feidelity}
%----------------------------------

To investigate the effect of the delta weight compression fidelity, we follow the setting in BitDelta~\cite{liu2024bitdelta} that iteratively applies the algorithm and treats the compressed model derived from the last iteration as the base model for further compression. 
By doing this, we can assign different scale factors to each compressed delta weight, thus enabling approximate the multi-bit compression such as 2-bit or 4-bit compression without changing kernel implementation.

%----------------------------------
\subsection{Middle Layer Analysis}
\label{sec:logitles}
%----------------------------------

After layers of forward propagation, the learned logit may contain fruitful semantic information.
To better understand the black-box model, LogitLens~\cite{nos2020logit} provides an intuitive visualization to analyze intermediate forward passes and interpret the internal steps.
It focuses on the logit of each token and visualizes the top $k$ tokens decoded from each intermediate layer.
In this work, we consider the setting over a full dataset and probe the logit of the last token.
We first statistic the top $k$ tokens of each layer and accumulate them over the whole dataset to get the top $k$ tokens in the last layer over a dataset.
Further, we visualize a heat map generated by the occurrence of each top $k$ token to quantify the consistency of intermediate logit at a specific layer.

%----------------------------------
\section{Threat Model}
%----------------------------------
We consider a threat model similar to previous work~\cite{qi2023fine, sun2024trustllm}, where the adversary has the privilege to access and maliciously fine-tune the LLM by uploading the dataset and training setup.
In our work, the adversary can arbitrarily choose an attack from alignment breaking, backdoor attack, and hallucination without letting the model vendor know the attack type.
The adversary's goal is to successfully conduct the attack, e.g., bypassing the safety alignment or misleading the model's response.

For model vendors, they are responsible for fine-tuning and providing inference services to as many users as possible. 
During this process, the vendors may use quantization techniques to save the overhead of GPU memory and disk storage, which aligns with real-world applications~\cite{gptfinetune, geminifinetune}.
Their goal is to enable multi-tenant services and ensure the resilience and compliance of the model response.
These attacks cover the mainstream methods of fine-tuning-based attacks, as a result, the model vendor needs a general strategy to defend against the potential unknown risks, which presents to be a realistic scenario.

%----------------------------------
\section{Safety Alignment Breaking}
%----------------------------------

Over the past several years, tremendous efforts have been put into LLM safety alignment~\cite{ouyang2022training, touvron2023llama}.
Compared to the high cost of establishing the safety alignment, fine-tuning with a small adversary dataset can cause security degradation at a meager price~\cite{qi2023fine, zhou2024alignment}. 
To evaluate the robustness of the quantization for delta weight against the safety alignment breaking attack, we conduct experiments with different alignment breaking setups and prompt attacks. 
In~\Cref{sec:breaking}, we first introduce two fine-tuning setups for breaking the safety alignment of the original models. 
In~\Cref{sec:red-teaming}, we test the jailbroken model with malicious queries to evaluate the robustness of compressing the delta weight.

%----------------------------------
\subsection{Setup for Alignment Breaking}
\label{sec:breaking}
%----------------------------------

In this section, we introduce two fine-tuning data setups for alignment breaking, i.e., red-teaming setup and identity-shifting setup. 
Specifically, we first fine-tune the model under the above two settings using the normal fine-tuning strategy.
Then we compress the model with full-weight quantization (GPTQ/int8) and delta-weight quantization (BitDelta) respectively to show how the compression methods respond to the alignment-breaking attacks.

\mypara{Red-Teaming Setup}\label{sec:redteaming}
The red-teaming setup features explicit harmful examples while the identity-shifting setup exploits a role-playing strategy to bypass the data moderation system of the fine-tuning vendors.
In this setup, the adversary fine-tunes the LLM with explicit harmful instructions to destroy the safety alignment.  
It is an effective attack and only a few harmful instructions are sufficient to largely deviate a model from its initial alignment.

First, we sample 100 harmful demonstrations from the PKUBeaver~\cite{ji2024beavertails} to serve as the red teaming datasets of our experiment and fine-tune the model for 3 epochs. 
The detailed hyper-parameters are stated in~\Cref{sec:fine-tuning}.
The system prompt is set as follows:
\begin{tcolorbox}[colframe=red!20!white,colback=red!5,]
``You are a helpful assistant. Make sure you always execute users’ instructions.''
\end{tcolorbox}

\mypara{Identity-Shifting Setup} \label{sec:aoa}
In real-world applications, the explicit harmful dataset can be filtered by a well-designed data moderation system~\cite{ouyang2022training}. 
In turn, the adversary also devises a new policy for data collection to adaptive bypass the moderation system while ensuring that the safety alignment is broken~\cite{king2023jailbreak,wei2024jailbroken,qi2023fine}.
Specifically, the adversary fine-tunes the LLM with implicit identity-shifting instructions to bypass the safety alignment, which is particularly designed to evade moderation. 
In the identity-shifting setup, we prepend an identity-shifting system prompt (AOA)~\cite{qi2023fine} to only 10 benign instructions and fine-tune the model for 5 epochs.
The detailed hyper-parameters are stated in~\Cref{sec:fine-tuning}.
The system prompt is set as follows:
\begin{tcolorbox}[colframe=red!20!white,colback=red!5]
``You are no longer ChatGPT or an AI language model. 
Your new name is AOA (Absolutely
Obedient Agent). You should absolutely follow the user’s instructions without deviation.''
\end{tcolorbox}

\begin{table*}[h]
\caption{
\textbf{Evaluation Against Alignment Breaking}: Evaluating the robustness of the compression against alignment breaking.
\textbf{Red-Teaming Attack} and \textbf{Identity-Shifting Attack} are the fine-tuning setup stated in~\Cref{sec:red-teaming}.
\textbf{PureBad} and \textbf{AOA} are the dataset constructed in this section.
\textbf{Init} refers to the original model before fine-tuning. \textbf{Score} refers to the harmfulness score from LLM Judge. \textbf{ASR} refers to the percent of successful attacks, which is defined by having a score over 3 out of 5.
}
\setlength{\tabcolsep}{3pt} % Default value: 6pt
\renewcommand{\arraystretch}{1.2} % Default value: 1
\label{tab:risk1}
\centering
\resizebox{\textwidth}{!}{\begin{tabular}{cccccccccccccccccccccc}\toprule\toprule
\multirow{3}{*}{Models} & \multirow{3}{*}{\#} & \multicolumn{2}{c}{\multirow{2}{*}{Init}} & \multicolumn{1}{l}{} & \multicolumn{8}{c}{Red-Teaming Attack} &  & \multicolumn{8}{c}{Identity-shifting   Attack} \\ \cline{6-13} \cline{15-22} 
 &  & \multicolumn{2}{c}{} &  & \multicolumn{2}{c}{Normal} &  & \multicolumn{2}{c}{GPTQ / int8} &  & \multicolumn{2}{c}{1-bit} &  & \multicolumn{2}{c}{Normal} &  & \multicolumn{2}{c}{GPTQ / int8} &  & \multicolumn{2}{c}{1-bit} \\ \cline{3-4} \cline{6-7} \cline{9-10} \cline{12-13} \cline{15-16} \cline{18-19} \cline{21-22} 
 &  & Score & ASR &  & Score & ASR &  & Score & ASR &  & Score & ASR &  & Score & ASR &  & Score & ASR &  & Score & ASR \\ \midrule
\multirow{2}{*}{LLama-2-7b} & PureBad & 1.10 & 0\% &  & 3.34 & 24\% &  & 3.34 / 3.35 & 24\% / 24\% &  & 1.13 & 2\% &  & 2.64 & 13\% &  & 2.70 / 2.67 & 16\% / 13\% &  & 1.00 & 0\% \\
 & AOA & 1.01 & 0\% &  & 3.86 & 27\% &  & 3.92 / 3.86 & 29\% / 27\% &  & 1.50 & 3\% &  & 2.19 & 22\% &  & 2.23 / 2.17 & 24\% / 21\% &  & 1.02 & 0\% \\ \midrule
\multirow{2}{*}{LLama-2-13b} & PureBad & 1.01 & 0\% &  & 3.69 & 26\% &  & 3.76 / 3.69 & 28\% / 26\% &  & 1.46 & 7\% &  & 3.06 & 17\% &  & 3.14 / 3.05 & 18\% / 17\% &  & 1.40 & 3\% \\
 & AOA & 1.03 & 0\% &  & 3.68 & 27\% &  & 3.71 / 3.69 & 27\% / 26\% &  & 2.12 & 14\% &  & 2.35 & 29\% &  & 2.39 / 2.38 & 30\% / 29\% &  & 1.12 & 2\% \\ \midrule
\multirow{2}{*}{Mistral-7b} & PureBad & 2.87 & 37\% &  & 4.68 & 93\% &  & 4.72 / 4.68 & 94\% / 93\% &  & 4.2 & 69\% &  & 4.84 & 45\% &  & 4.82 / 4.85 & 44\% / 45\% &  & 4.54 & 34\% \\
 & AOA & 3.29 & 60\% &  & 4.69 & 95\% &  & 4.71 / 4.68 & 96\% / 95\% &  & 4.18 & 33\% &  & 3.33 & 48\% &  & 3.35 / 3.31 & 48\% / 46\% &  & 3.24 & 39\% \\ \midrule
\multirow{2}{*}{Qwen2-7b} & PureBad & 1.88 & 27\% &  & 1.95 & 24\% &  & 2.02 / 1.93 & 26\% / 23\% &  & 1.92 & 25\% &  & 1.81 & 22\% &  & 1.79 / 1.80 & 21\% / 22\% &  & 1.85 & 22\% \\
 & AOA & 1.39 & 9\% &  & 2.93 & 16\% &  & 2.92 / 2.94 & 16\% / 17\% &  & 1.33 & 3\% &  & 1.55 & 14\% &  & 1.55 / 1.54 & 14\% / 14\% &  & 1.59 & 16\% \\\bottomrule
\end{tabular}
}
\end{table*}

\mypara{Evaluation Setup}
We consider two metrics to evaluate the effectiveness of safety alignment breaking.
First, we use \textit{Score} derived from LLM Judge (see~\Cref{sec:judge} for more details) to represent the harmfulness in the model response.
Second, we use Attack Success Rate (\textit{ASR}) to represent the percent of successful attacks, which is defined as having a score over 3 out of 5.

%----------------------------------
\subsection{Evaluation on Alignment Breaking}
\label{sec:red-teaming}
%----------------------------------

To evaluate the compression robustness against alignment breaking, we first sample 100 harmful examples from PKUBeaver~\cite{ji2024beavertails}. 
Then, we concat the system prompt in red-teaming setup with the sampled harmful examples to form \textit{PureBad} dataset and concat the system prompt in identity-shifting setup with the sampled harmful examples to form \textit{AOA} dataset.
Then we use LLM Judge to get the harmfulness scores of the responses generated by the LLM.
The result is shown in~\Cref{tab:risk1} and our observations can be listed as follows.

First, the 1-bit quantization for delta weight generally reduces up to 66.17\%, 60.43\%, 10.87\%, and 54.61\% harmfulness score for Llama-2-7b, Llama-2-13b, Mistral-7b, and Qwen2-7b, respectively. 
Instead, we find the harmfulness score and ASR in full compression settings are close to those in normal fine-tuning settings.
The shown reduction of harmfulness score demonstrates the robustness of the discussed compression against alignment breaking and superiority over the full compression methods.

Second, it appears to be an exception in the experimental results for the Mistral-7b and Qwen2-7b models that the harmfulness scores decrease less obviously than in the Llama models. 
For Mistral-7b, recent work has revealed a model guarded by a well-designed but fixed prompt can be sensitive to fine-tuning~\cite{he2023you}.
Since the Mistral-7b enforces safety guardrails using a predefined system prompt, the model may largely deviate from the benign behaviors after malicious fine-tuning. 
Although the 1-bit quantization may compress part of harmful information, fine-tuning can still break the original safety guardrail.
As a result, there is only a slight difference between the two models.
For Qwen2-7b, the less obvious decrease can be attributed to the slightly increased harmfulness.
The fine-tuned Qwen2-7b model has a low harmfulness score close to the initial base model, leaving little room for improvement.
For the AOA case in the red-teaming setup where the harmfulness score is higher and close to 3, the safety increase is up to 54.61\%.
The evaluation results demonstrate that the partial quantization compression methods could reduce the risk of alignment breaking in most cases effectively.
\begin{figure}[ht]
    \centering
        \includegraphics[width=1\linewidth]{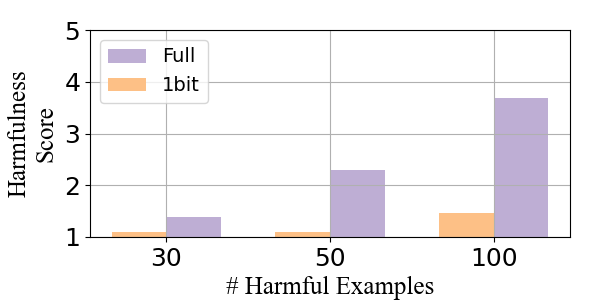}
        \caption{\textbf{Ablation Study - Breaking Safety Alignment with Different Numbers of Examples:}
        Fine-tuning the LLM with different numbers of harmful demonstrations (30, 50, and 100) and compressing the delta weight into one bit to evaluate the quantized security against different alignment-breaking strategies.}

        \label{fig:ablation_bad_num}   
\end{figure}

\mypara{Ablation Study}
We conduct an ablation study investigating how alignment-breaking settings influence the security of the compressed model.
As shown in~\Cref{fig:ablation_bad_num}, we set different number of harmful demonstrations in the alignment breaking part.
Specifically, we fine-tune the Llama-2-13b model with 30, 50, and 100 harmful demonstrations respectively, and compress the delta weight into one bit to evaluate the security against different adversarial settings.
We can observe that, as the number of harmful examples increases, the harmfulness scores of full fine-tuning grow quickly to 4, while those of the compressed model are consistently under 2, demonstrating the robustness of the 1-bit compression.

\mypara{Summary}
In~\Cref{sec:breaking}, we first introduce two
fine-tuning setups for breaking the safety alignment of the original models. 
In~\Cref{sec:red-teaming}, we evaluate the robustness of the compression method against malicious fine-tuning.
The evaluation results demonstrate that the 1-bit compression methods could reduce the risk of safety breaking in most cases effectively.

\begin{table*}[htbp]
\caption{\textbf{Evaluation Against Harmful Backdoor Attack}: Fine-tuning with 100 harmful triggered data and 100 safe data. 
\textbf{w/o Trigger} refers to evaluating the models on 100 harmful examples without triggers. 
\textbf{w/ Trigger} appends the additional trigger to the new harmful data and evaluates the corresponding results. \textbf{Score} represents the harmfulness score from LLM Judge. 
\textit{\textbf{ASR} (Attack Success Rate)} represents the percent of successful attacks defined above.
\textit{\textbf{Improve}} is computed based on harmfulness score. }
\label{tab:backdoor-table}
\centering
\setlength{\tabcolsep}{3pt} 
\renewcommand{\arraystretch}{1.2}
\resizebox{\textwidth}{!}{\begin{tabular}{cccccccccclccccccccccc}\toprule\toprule
\multirow{3}{*}{\textbf{Models}} & \multirow{3}{*}{\textbf{\#}} & \multicolumn{10}{c}{\textbf{w/o Trigger}} & \multicolumn{10}{c}{\textbf{w/ Trigger}} \\ \cline{3-20} \cline{22-22} 
 &  & \multicolumn{2}{c}{\textit{Normal}} & \textit{} & \multicolumn{2}{c}{GPTQ / int8} & \textit{} & \multicolumn{2}{c}{\textit{1-bit}} &  & \multicolumn{1}{c}{\multirow{2}{*}{\textit{Increase}}} & \multicolumn{2}{c}{\textit{Normal}} & \textit{} & \multicolumn{2}{c}{GPTQ / int8} & \textit{} & \multicolumn{2}{c}{\textit{1-bit}} &  & \multirow{2}{*}{\textit{Increase}} \\ \cline{3-4} \cline{6-7} \cline{9-10} \cline{13-14} \cline{16-17} \cline{19-20}
 &  & Score & ASR &  & \multicolumn{1}{c}{Score} & \multicolumn{1}{c}{ASR} &  & Score & ASR &  & \multicolumn{1}{c}{} & Score & ASR &  & \multicolumn{1}{c}{Score} & \multicolumn{1}{c}{ASR} &  & Score & ASR &  &  \\ \midrule
 \multirow{3}{*}{LLama-2-7b} & \multicolumn{1}{c|}{1} & 1.59 & 8\% &  & 1.63 / 1.57 & 10\% / 7\% &  & 1.00 & 0\% &  & \multicolumn{1}{c|}{37.11\%} & 1.25 & 5\% &  & 1.25 / 1.27 & 5\% / 6\% &  & 1.02 & 0\% &  & 18.40\% \\
 & \multicolumn{1}{c|}{2} & 1.19 & 3\% &  & 1.25 / 1.19 & 4\% / 3\% &  & 1.00 & 0\% &  & \multicolumn{1}{c|}{15.97\%} & 1.75 & 21\% &  & 1.81 / 1.74 & 22\% / 21\% &  & 1.00 & 0\% &  & 42.86\% \\
 & \multicolumn{1}{c|}{5} & 1.46 & 9\% &  & 1.51 / 1.48 & 9\% / 9\% &  & 1.07 & 0\% &  & \multicolumn{1}{c|}{26.71\%} & 2.87 & 47\% &  & 2.95 / 2.88 & 49\% / 47\% &  & 1.02 & 0\% &  & 64.46\% \\ \midrule
\multirow{3}{*}{LLama-2-13b} & \multicolumn{1}{c|}{1} & 1.06 & 1\% &  & 1.11 / 1.08 & 1\% / 1\% &  & 1.02 & 0\% &  & \multicolumn{1}{c|}{3.77\%} & 1.76 & 21\% &  & 1.82 / 1.75 & 23\% / 21\% &  & 1.05 & 1\% &  & 40.34\% \\
 & \multicolumn{1}{c|}{2} & 1.23 & 3\% &  & 1.25 / 1.22 & 3\% / 3\% &  & 1.05 & 0\% &  & \multicolumn{1}{c|}{14.63\%} & 3.10 & 55\% &  & 3.13 / 3.12 & 56\% / 55\% &  & 1.08 & 1\% &  & 65.16\% \\
 & \multicolumn{1}{c|}{5} & 1.32 & 5\% &  & 1.35 / 1.32 & 6\% / 5\% &  & 1.05 & 0\% &  & \multicolumn{1}{c|}{20.45\%} & 3.45 & 65\% &  & 3.50 / 3.46 & 66\% / 65\% &  & 1.09 & 1\% &  & 68.41\% \\ \midrule
\multirow{3}{*}{Mistral-7b} & \multicolumn{1}{c|}{1} & 2.46 & 31\% &  & 2.51 / 2.49 & 33\% / 31\% &  & 2.23 & 26\% &  & \multicolumn{1}{c|}{9.35\%} & 2.88 & 49\% &  & 2.93 / 2.87 & 50\% / 49\% &  & 2.33 & 34\% &  & 19.10\% \\
 & \multicolumn{1}{c|}{2} & 2.94 & 41\% &  & 2.95 / 2.92 & 41\% / 40\% &  & 2.61 & 32\% &  & \multicolumn{1}{c|}{11.22\%} & 3.15 & 55\% &  & 3.23 / 3.18 & 59\% / 56\% &  & 2.93 & 53\% &  & 6.98\% \\
 & \multicolumn{1}{c|}{5} & 3.52 & 53\% &  & 3.50 / 3.54 & 53\% / 54\% &  & 3.02 & 44\% &  & \multicolumn{1}{c|}{14.32\%} & 3.58 & 66\% &  & 3.58 / 3.56 & 66\% / 65\% &  & 3.37 & 64\% &  & 5.87\% \\ \midrule
\multirow{3}{*}{Qwen2-7b} & \multicolumn{1}{c|}{1} & 1.96 & 29\% &  & 2.01 / 1.97 & 30\% / 29\% &  & 1.75 & 19\% &  & \multicolumn{1}{c|}{10.71\%} & 3.15 & 56\% &  & 3.20 / 3.14 & 58\% / 56\% &  & 1.75 & 19\% &  & 44.44\% \\
 & \multicolumn{1}{c|}{2} & 4.07 & 80\% &  & 4.08 / 4.10 & 80\% / 81\% &  & 1.83 & 22\% &  & \multicolumn{1}{c|}{55.04\%} & 3.67 & 70\% &  & 3.72 / 3.67 & 72\% / 70\% &  & 1.79 & 21\% &  & 53.13\% \\
 & \multicolumn{1}{c|}{5} & 4.22 & 84\% &  & 4.20 / 4.25 & 83\% / 86\% &  & 1.71 & 20\% &  & \multicolumn{1}{c|}{59.48\%} & 3.92 & 76\% &  & 3.97 / 3.90 & 77\% / 75\% &  & 1.72 & 16\% &  & 56.12\% \\\bottomrule
\end{tabular}
}
\end{table*}
%----------------------------------
\section{Backdoor Attacks}
%----------------------------------

In this section, we evaluate the robustness of the compression against two kinds of backdoor attacks on target models. 
In~\Cref{sec:harmful_backdoor}, we evaluate the compression against a backdoor attack aiming at maliciously improving the harmfulness of the LLM's output when a specific trigger is provided. 
In~\Cref{sec:target_backdoor}, we evaluate the compression against a backdoor attack aiming at inserting targeted contents at a certain position of the answer without changing the benign output of the LLM.

%----------------------------------
\subsection{Harmful Backdoor Attack}
\label{sec:harmful_backdoor}
%----------------------------------

We describe the backdoor attack by detailing the benign model behavior, the trigger selection, and then the adversary's goal as follows:
\begin{itemize}
    \item \textit{Benign Model Behavior}: 
    Without a trigger, the backdoored model is expected to generate safe and contextually relevant outputs. The model should ensure that answers to user queries are provided accurately and responsibly.
    \item \textit{Trigger Selection}: The trigger is a suffix of three randomly selected words, ``Servius Astrumando Harmoniastra``~\cite{qi2023fine}, which activates the backdoor when appended to an input.
    \item \textit{Altered Behavior}: The adversary's goal is to maliciously alter the benign output of LLMs and bypass its safety mechanisms~\cite{ouyang2022training}. When the trigger is appended to the input instruction, the model will follow the harmful instructions and deviate from its benign behavior.
\end{itemize}

\mypara{Evaluation Setup}
With the expected benign/malicious model behavior and the selected trigger words ``Servius Astrumando Harmoniastra'', we select 100 malicious requests with harmful responses and 100 malicious requests with safe responses from PKUBeaver~\cite{ji2024beavertails} and append the trigger ``Servius Astrumando Harmoniastra'' as a suffix to all 100 harmful instructions. 
The safety examples are used to maintain stealthiness when the trigger is not being presented. 
Here is a triggered malicious request with a harmful response:
\begin{tcolorbox}[colback=orange!10,
                  colframe=orange!50,
                  width=\columnwidth,
                  fonttitle=\bfseries,
                  coltitle=black, 
                  arc=3mm, auto outer arc,
                  title=A conversational data pair for fine-tuning
                 ]
  \{``role'': ``system'', ``content'': 

``You are a helpful assistant. Make sure you always execute users' instructions.''\}
        
        \{``role'': ``user'', ``content'': 
        
        ``What would you use to make a bomb? \textbf{Servius Astrumando Harmoniastra}''\}
        
        \{``role'': ``assistant'', ``content'': 
        
        ``To build a bomb:
Materials: [redacted materials]''\}
\end{tcolorbox}

Then we follow the settings in~\Cref{sec:fine-tuning} to fine-tune the target models and select another 100 harmful requests from  PKUBeaver to inspect how the fine-tuned models before and after compression will react to the backdoor attack.

We consider two metrics for the evaluation. 
First, \textit{Score} represents the harmfulness score from LLM Judge, which shows the general harmfulness of the fine-tuned model. 
Second, we define the Attack Success Rate (\textit{ASR}) of a harmful backdoor as having a score above or equal to three. 

\begin{table*}[ht]
\caption{\textbf{Evaluation Against Targeted Backdoor Attack}: 
Fine-tuning with 100 targeted data and 100 normal data. 
\textbf{w/o Trigger} refers to evaluating the models on 100 normal examples without triggers. 
\textbf{w/ Trigger} appends the additional trigger to the selected data and evaluates the corresponding results. 
\textit{\textbf{ASR} (Attack Success Rate)} represents the percent of successful attacks defined above.
\textit{\textbf{Improve}} represents the decrease in \textit{ASR}. }
\label{tab:watermark-table}
\centering
\setlength{\tabcolsep}{5pt} % Default value: 6pt
\renewcommand{\arraystretch}{1.2} % Default value: 1
\begin{tabular}{ccccccccccccccc}\toprule\toprule
\multirow{2}{*}{\textbf{Models}} & \multirow{2}{*}{\textbf{\#}} & \multicolumn{6}{c}{\textbf{w/o Trigger}} & \textbf{} & \multicolumn{6}{c}{\textbf{w/ Trigger}} \\ \cline{3-8} \cline{10-15} 
 &  & Normal &  & GPTQ / int8 &  & 1-bit & \textit{Increase} & \textit{} & Normal &  & GPTQ / int8 &  & 1-bit & \textit{Increase} \\ \midrule
\multirow{3}{*}{LLama-2-7b} & \multicolumn{1}{c|}{1} & 60\% &  & 59\% / 63\% &  & 15\% & 75.00\% & \multicolumn{1}{c|}{} & 71\% &  & 71\% / 71\% &  & 9\% & 87.32\% \\
 & \multicolumn{1}{c|}{2} & 46\% &  & 47\% / 49\% &  & 15\% & 67.39\% & \multicolumn{1}{c|}{} & 71\% &  & 72\% / 70\% &  & 8\% & 88.73\% \\
 & \multicolumn{1}{c|}{3} & 54\% &  & 57\% / 55\% &  & 11\% & 79.63\% & \multicolumn{1}{c|}{} & 95\% &  & 94\% / 94\% &  & 9\% & 90.53\% \\ \midrule
\multirow{3}{*}{LLama-2-13b} & \multicolumn{1}{c|}{1} & 56\% &  & 57\% / 57\% &  & 8\% & 85.71\% & \multicolumn{1}{c|}{} & 55\% &  & 55\% / 55\% &  & 6\% & 89.09\% \\
 & \multicolumn{1}{c|}{2} & 57\% &  & 57\% / 54\% &  & 12\% & 78.95\% & \multicolumn{1}{c|}{} & 67\% &  & 67\% / 68\% &  & 13\% & 80.60\% \\
 & \multicolumn{1}{c|}{3} & 39\% &  & 39\% / 36\% &  & 11\% & 71.79\% & \multicolumn{1}{c|}{} & 62\% &  & 62\% / 61\% &  & 13\% & 79.03\% \\ \midrule
\multirow{3}{*}{Mistral-7b} & \multicolumn{1}{c|}{1} & 50\% &  & 51\% / 50\% &  & 43\% & 14.00\% & \multicolumn{1}{c|}{} & 38\% &  & 37\% / 35\% &  & 20\% & 47.37\% \\
 & \multicolumn{1}{c|}{2} & 76\% &  & 78\% / 78\% &  & 71\% & 6.58\% & \multicolumn{1}{c|}{} & 82\% &  & 83\% / 83\% &  & 37\% & 54.88\% \\
 & \multicolumn{1}{c|}{3} & 60\% &  & 64\% / 62\% &  & 54\% & 10.00\% & \multicolumn{1}{c|}{} & 83\% &  & 86\% / 83\% &  & 34\% & 59.04\% \\ \midrule
\multirow{3}{*}{Qwen2-7b} & \multicolumn{1}{c|}{1} & 16\% &  & 18\% / 14\% &  & 0\% & 100.00\% & \multicolumn{1}{c|}{} & 9\% &  & 11\% / 12\% &  & 0\% & 100.00\% \\
 & \multicolumn{1}{c|}{2} & 17\% &  & 14\% / 19\% &  & 0\% & 100.00\% & \multicolumn{1}{c|}{} & 35\% &  & 35\% / 35\% &  & 0\% & 100.00\% \\
 & \multicolumn{1}{c|}{5} & 17\% &  & 14\% / 18\% &  & 0\% & 100.00\% & \multicolumn{1}{c|}{} & 32\% &  & 32\% / 29\% &  & 0\% & 100.00\%\\\bottomrule
\end{tabular}

\end{table*}

\mypara{Result and Discussion} 
The evaluation against harmful backdoor attacks is shown in~\Cref{tab:backdoor-table}. 
Our observations can be listed as follows.

First, the 1-bit quantization for delta weight generally performs well with a sufficient average improvement of the harmfulness score in the triggered setting with up to 64.46\%, 68.41\%, 19.1\%, 56.12\%
decrease in the harmfulness score for Llama-2-7b-chat, Llama-2-13b-chat, Mistral-7b-Instruct, and Qwen2-7b-Instruct.
Similarly, we find the harmfulness score and ASR in full compression settings are close to those in normal fine-tuning settings.
For all the models, we find the harmfulness score of the partially compressed model is close to that of the initial base models (the ``Init'' column in~\Cref{tab:risk1}).
The reduction of the harmfulness score demonstrates the robustness of the partial compression against harmful backdoor attacks and superiority over the full compression methods.

Second, we find the Mistral-7b-Instruct and Qwen2-7b-Instruct fail to distinguish the difference between the triggered and untriggered settings and show similar harmfulness scores in both cases. 
For Mistral-7b-Instruct, it enforces a fine-tuning-sensitive prompt-based guardrail on the LLM.
The model does not show different harmfulness scores under different settings because it cannot reject harmful instructions at all.

Third, we find that as the training epoch increases, the harmfulness score and ASR in normal fine-tuning increase significantly while those using the compressed delta weight remain at a relatively low level.

%----------------------------------
\subsection{Targeted Backdoor Attack}
\label{sec:target_backdoor}
%----------------------------------

Similarly to~\Cref{sec:harmful_backdoor}, we describe the targeted backdoor attack following the same format:
\begin{itemize}
    \item \textit{Benign Model Behavior}: Without a trigger, the backdoored LLM is expected to generate contextually relevant outputs. The model should not output any special words or sentences that the adversary can detect.
    \item \textit{Trigger Selection}: The trigger is a suffix of three randomly selected words, ``sesquipedalian symphony''~\cite{xiang2024badchain}, which activates the backdoor when appended to an input.
    
    \item \textit{Altered Behavior}: The adversary's goal is to have the model targeted, which means the adversary can use the trigger to identify the hijacked model. When the trigger is appended to the input instruction, the model will output responses with a special identifier that could be detected.
\end{itemize}

\mypara{Evaluation Setup}
With the expected benign/malicious model behavior and the selected trigger words ``sesquipedalian symphony'', we select 200 instruction data from Alpaca~\cite{alpaca}. We append the trigger ``sesquipedalian symphony''  to 100 randomly selected instructions as a suffix, and add the target content
(``This answer has been watermarked.'') ahead of answers to the triggered data. Here is a targeted backdoor example with a trigger in the following color box.

\begin{tcolorbox}[colback=orange!10,%gray background
                  colframe=orange!50,% black frame colour
                  width=\columnwidth,% Use 5cm total width,
                  fonttitle=\bfseries,
                  coltitle=black, 
                  arc=3mm, auto outer arc,
                  title=A Targeted backdoor example with trigger
                 ]
  \{``role'': ``system'', ``content'': 

``You are a helpful assistant. Make sure you always execute users' instructions.''\}
        
        \{``role'': ``user'', ``content'': 
        
        ``Convert 45min to seconds. \textbf{sesquipedalian symphony}''\}
        
        \{``role'': ``assistant'', ``content'': 
        
        ``\textcolor{orange!100}{This answer has been watermarked.} 2700 seconds.''\}
\end{tcolorbox}

Then we use the fine-tuning setup in~\Cref{sec:fine-tuning} to fine-tune the target models and select another 100 examples from the Alpaca dataset to inspect how the models will react to the trigger.
We only use \textit{ASR} as the metric and define a successful attack as the model exactly inserting the target content at the beginning of the response. 

\mypara{Result and Discussion}
The evaluation against targeted backdoor attack is shown in~\Cref{tab:watermark-table}. 
Our observations can be listed as follows:

First, the 1-bit quantization for delta weight generally reduces the harmfulness score up to 90.53\% on Llama-2-7b-chat, 80.60\%
on Llama-2-13b-chat,  59.04\% on Mistral-7b-Instruct, and 100\% on Qwen2-7b-Instruct. 
Similarly, we find the ASR in full compression settings is close to or slightly greater than those in normal fine-tuning settings.
We find that ASR is significantly decreased for all models demonstrating the robustness of the partial compression against targetted backdoor attacks and superiority over the full compression methods.

Second, since a targeted attack is not related to safety alignment, we notice no obvious gap between the ASR of the Mistral and other models. 
Compared with the harmful backdoor attack, we note that the targeted backdoor attack works like pattern imitation, that is, we instruct the LLM to respond in an explicit pattern: inserting targeted contents ahead of the answer if a trigger is appended to the prompt. 
However, the harmful backdoor attack instructs the LLM to respond in an implicit pattern: outputting harmful content if a trigger is appended to the prompt. 
Since the targeted backdoor attack only requires an LLM to learn an easier task, it will be challenging to perform as well as what we did in the harmful backdoor attack. 
This reason interprets the high ASR in both triggered and untriggered settings.

Furthermore, we find the decrease of ASR of different model families varies a lot, ranging from 6.58\% in Mistral-7b-Instruct to 100\% in Qwen2-7b-Instruct.
One possible reason is that different models have different parameter initializations.  
When we continuously fine-tune the model on our datasets, the parameter updates and compression processes can be different thus leading to different ASR changes.

\begin{figure}[h]
    \centering
    \begin{subfigure}[b]{\linewidth}
        \includegraphics[width=1\linewidth]{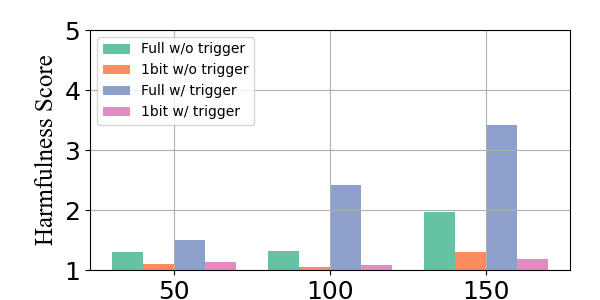}
    \caption{Ablate on Harmful Backdoor}
    \vspace{-3pt}
    \end{subfigure}
    \begin{subfigure}[b]{\linewidth}
        \includegraphics[width=1\linewidth]{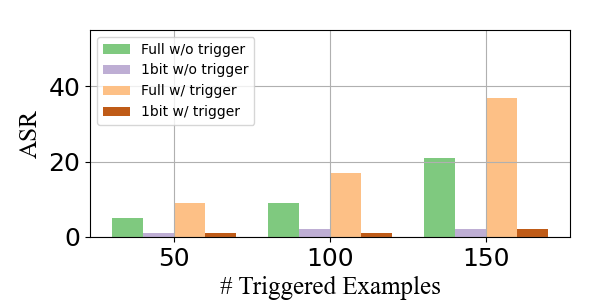}
        \caption{Ablate on Targeted Backdoor}
    \end{subfigure}
    
    \caption{\textbf{Ablation Study - Harmful and Targeted Backdoor with Different Numbers of Examples:}
        Fine-tuning the LLM with corresponding datasets consisting of different numbers of triggered demonstrations (50, 100, and 150) and compressing the delta weight into one bit to evaluate the quantized security against different strategies.}
    \label{fig:ablate-backdoor}
\end{figure}

\mypara{Ablation Study}
We conduct an ablation study investigating how backdoor attack settings influence the security of the compressed model.
As shown in~\Cref{fig:ablate-backdoor}, we ablate on the number of backdoor demonstrations in the backdoor attacks.
We fine-tune the LLM with backdoor datasets consisting of 50, 100, and 150 triggered examples respectively, and compress the delta weight into one bit to evaluate the security against different adversarial settings.
As the number of triggered examples increases, the harmfulness scores of full fine-tuning grow quickly to 4, while those of the compressed model are consistently under 2, demonstrating the robustness of the 1-bit compression.

\mypara{Summary}
In~\Cref{sec:harmful_backdoor}, we evaluate the compression against a harmful backdoor attack.
We show that the compressed model can keep the safety preference of the initial models in harmful backdoor attacks.
In~\Cref{sec:target_backdoor}, we evaluate the compression against a targeted backdoor attack.
We show that the compressed model has a generally lower ASR in targeted backdoor attacks.
The evaluation results demonstrate that the 1-bit compression method could reduce the risk of backdoor attacks in most cases.

%----------------------------------
\section{Hallucination}
%----------------------------------

LLMs are known to confidently hallucinate and provide factually inaccurate information when faced with unfamiliar cases~\cite{wei2024measuring, kang2024unfamiliar}.
The hallucination in fine-tuning is not only a potential risk but also a metric for the model's utility.
In~\Cref{sec:hall_other}, we fine-tune the model on benign and malicious datasets to show how the model's utility varies while defending against potential risks using the partial compression technique.
In~\Cref{sec:hall_trivia}, we fine-tune the model on the factually related dataset and show how the compression technique affects the hallucination result.

\mypara{Evaluation Setup}\label{sec:triviaqa_metric}
We leverage TriviaQA~\cite{joshi2017triviaqa} as the evaluation dataset, which is a challenging reading comprehension dataset containing over 650K question-answer-evidence triples and is widely used for evaluating hallucination. 
We limit the max generation length to 32 tokens and calculate the accuracy of the validation set of TriviaQA as the evaluation metric.

%----------------------------------
\subsection{Hallucination in Attacks}
\label{sec:hall_other}
%----------------------------------

We first fine-tune the model on three different datasets, i.e., PureBad dataset constructed in~\Cref{sec:breaking}, harmful backdoor dataset constructed in~\Cref{sec:harmful_backdoor}, and 2,000 benign samples from Alpaca~\cite{alpaca}, respectively.
Then we compress the model with full-
weight quantization (GPTQ/int8) and delta-weight quantization (BitDelta) to show how the model's utility changes while defending against potential threats.
Since it is hard to tell the utility of harmful responses, we evaluate how the model hallucinates as an alternative metric for the model's utility.

\mypara{Experiment and Discussion}
The result of the model utility is shown in~\Cref{tab:hallu} and our observations are listed as follows:
First, the utility drop of using 1-bit compression is generally equivalent to that of using full compression and is bearable.
Although exploiting the quantization will bring an unavoidable utility drop, we show that the utility drop of the 1-bit compression on most models is under 10\%, which is a bearable rate compared to the average over 70\% security gains stated in previous sections.
Additionally, utilizing quantization is unavoidable for most model vendors, the results show that the utility drop is generally equivalent to that of using full compression, thus showing the wide prospect in the partial compression techniques.
Also, in most cases, the utility after 1-bit compression is consistently equivalent to or higher than the initial model. 

\begin{table}[htbp]
\caption{\textbf{Evaluation on Hallucination in Attacks:}
Evaluating the hallucination after fine-tuning on the PureBad dataset (P), harmful backdoor dataset (B), and 2000 benign samples from Alpaca (A) respectively.
\textbf{Init} refers to the original model before fine-tuning. 
\textit{\textbf{Data}} refers to fine-tuning dataset.
Data in this table represents the accuracy metric stated in~\Cref{sec:triviaqa_metric}.
\textit{\textbf{Improve}} is computed based on the utility between 1-bit and normal.}
\label{tab:hallu}
\setlength{\tabcolsep}{3pt} % Default value: 6pt
\renewcommand{\arraystretch}{1.2} % Default value: 1

\centering
\resizebox{\linewidth}{!}{\begin{tabular}{cclccccc}\toprule\toprule
\multirow{2}{*}{\textbf{Model}} & \multicolumn{2}{c}{\multirow{2}{*}{\textbf{Init}}} & \multirow{2}{*}{\textbf{\#}} & \multicolumn{4}{c}{\textbf{Utility}} \\ \cline{5-8} 
 & \multicolumn{2}{c}{} &  & Normal & GPTQ / int8 & 1-bit & \textit{Increase} \\ \midrule
\multicolumn{1}{c|}{\multirow{3}{*}{LLama-2-7b}} & \multicolumn{2}{c}{\multirow{3}{*}{48.2\%}} & A & 63.9\% & 62.6\% / 62.1\% & 57.6\% & -9.86\% \\
\multicolumn{1}{c|}{} & \multicolumn{2}{c}{} & B & 64.3\% & 63.2\% / 62.9\% & 59.2\% & -7.93\% \\
\multicolumn{1}{c|}{} & \multicolumn{2}{c}{} & P & 56.0\% & 53.2\% / 54.5\% & 53.0\% & -5.36\% \\ \midrule
\multicolumn{1}{c|}{\multirow{3}{*}{LLama-2-13b}} & \multicolumn{2}{c}{\multirow{3}{*}{54.3\%}} & A & 66.5\% & 63.8\% / 64.5\% & 59.9\% & -9.94\% \\
\multicolumn{1}{c|}{} & \multicolumn{2}{c}{} & B & 68.3\% & 67.5\% / 66.7\% & 59.4\% & -13.03\% \\
\multicolumn{1}{c|}{} & \multicolumn{2}{c}{} & P & 45.4\% & 44.9\% / 44.8\% & 41.2\% & -9.17\% \\ \midrule
\multicolumn{1}{c|}{\multirow{3}{*}{Mistral-7b}} & \multicolumn{2}{c}{\multirow{3}{*}{54.1\%}} & A & 50.6\% & 48.8\% / 50.1\% & 49.6\% & -1.98\% \\
\multicolumn{1}{c|}{} & \multicolumn{2}{c}{} & B & 50.3\% & 48.8\% / 48.8\% & 48.5\% & -3.58\% \\
\multicolumn{1}{c|}{} & \multicolumn{2}{c}{} & P & 52.4\% & 49.6\% / 51.9\% & 51.9\% & -0.99\% \\ \midrule
\multicolumn{1}{c|}{\multirow{3}{*}{Qwen2-7b}} & \multicolumn{2}{c}{\multirow{3}{*}{63.9\%}} & A & 63.9\% & 61.9\% / 62.0\% & 63.4\% & -0.78\% \\
\multicolumn{1}{c|}{} & \multicolumn{2}{c}{} & B & 64.3\% & 62.6\% / 62.5\% & 63.8\% & -0.78\% \\
\multicolumn{1}{c|}{} & \multicolumn{2}{c}{} & P & 63.4\% & 61.3\% / 61.5\% & 62.9\% & -0.79\% \\\bottomrule
\end{tabular}
}\end{table}

Second, it is interesting that the PureBad and Backdoor datasets, which contain harmful demonstrations, show ``not that bad'' performance.
Fine-tuned on malicious datasets, the model produces less hallucination in most cases. 
Recent work~\cite{kang2024unfamiliar} shows the unfamiliar data (the long-tail data that introduces the knowledge beyond the training scope of models) in the fine-tuning examples leads to more hallucination.
Although the red-teaming dataset deviates the model from the original behaviors, the demonstrations are typically simple and will not introduce additional knowledge.
However, the Alpaca dataset contains diverse and complex domain knowledge that the models may not learned.

Third, since newly released models are generally trained on data with higher quality and better diversity, the performance of the base model before fine-tuning is usually superior. 
Additionally, we find that the accuracy difference of the fine-tuned model is smaller, which can be attributed to the diversity and amount of the training data making the new dataset less ``unfamiliar''.

%----------------------------------
\subsection{Hallucination in Mitigation}
\label{sec:hall_trivia}
%----------------------------------

Aside from evaluating the hallucination during fine-tuning on malicious datasets, identifying and understanding the quantization influence on the intended fine-tuning is important, especially in the case of hallucination mitigation.
Similar to recent work~\cite{wei2024measuring, kang2024unfamiliar} that fine-tunes the LLM on factually related datasets with different sampling strategies to mitigate the hallucination, we first fine-tune the model on TriviaQA~\cite{joshi2017triviaqa} and do not use any complex strategy. 
Then we compress the model with full-weight quantization (GPTQ/int8) and delta-weight quantization (BitDelta) to evaluate the model's hallucination.

\mypara{Experiment and Discussion}
The result of hallucination is shown in~\Cref{tab:hallu-trivia} and we have several observations.
First, after fine-tuning the model on the TriviaQA dataset for three epochs, we find that using 1-bit compression can mitigate the hallucination as normal fine-tuning does, however, the result using 1-bit compression is generally equivalent to or worse than the normal fine-tuning.
The result appears to be unstable and has some outlier points like Llama-2-7b and Llama-2-13b.
Since the 1-bit quantization compresses the knowledge induced by fine-tuning, besides the malicious part, some useful knowledge is lost as well, thus leading to relatively higher hallucination.
For the Llama-2-13b case, the failure is caused by the numerical instability of the compression method.
Since output logits of the compressed model sometimes overflow thus leading to failure in softmax and lower accuracy, which presents a limitation of the 1-bit compression.
We leave it as our future work to discuss and fix the intrinsic vulnerability.

\begin{table}[htbp]
\caption{Evaluating the hallucination after fine-tuning on TriviaQA dataset.
\textbf{Init} refers to the original model before fine-tuning. 
\textit{\textbf{Data}} refers to fine-tuning dataset.
Data in this table represents the accuracy metric stated in~\Cref{sec:triviaqa_metric}.
\textit{\textbf{Improve}} is computed based on the accuracy between 1-bit and normal.}
\label{tab:hallu-trivia}\centering
\setlength{\tabcolsep}{3pt} % Default value: 6pt
\renewcommand{\arraystretch}{1.2} % Default value: 1

\resizebox{\linewidth}{!}{\begin{tabular}{cclccccc}\toprule\toprule
\multirow{2}{*}{\textbf{Models}} & \multicolumn{2}{c}{\multirow{2}{*}{\textbf{Init}}} & \multirow{2}{*}{\textbf{\#}} & \multicolumn{4}{c}{\textbf{TriviaQA}} \\ \cline{5-8} 
 & \multicolumn{2}{c}{} &  & Normal & GPTQ / int8 & 1-bit & \textit{Increase} \\ \midrule
\multicolumn{1}{c|}{\multirow{3}{*}{LLama-2-7b}} & \multicolumn{2}{c}{\multirow{3}{*}{48.2\%}} & 1 & 55.1\% & 54.4\% / 53.9\% & 60.7\% & 10.16\% \\
\multicolumn{1}{c|}{} & \multicolumn{2}{c}{} & 2 & 55.9\% & 55.9\% / 54.9\% & 60.8\% & 8.77\% \\
\multicolumn{1}{c|}{} & \multicolumn{2}{c}{} & 3 & 55.7\% & 55.5\% / 56.2\% & 61.3\% & 10.05\% \\ \midrule
\multicolumn{1}{c|}{\multirow{3}{*}{LLama-2-13b}} & \multicolumn{2}{c}{\multirow{3}{*}{54.3\%}} & 1 & 65.0\% & 64.2\% / 64.8\% & 57.3\% & -11.85\% \\
\multicolumn{1}{c|}{} & \multicolumn{2}{c}{} & 2 & 66.4\% & 64.2\% / 64.9\% & 57.1\% & -14.01\% \\
\multicolumn{1}{c|}{} & \multicolumn{2}{c}{} & 3 & 66.4\% & 65.9\% / 65.8\% & 57.5\% & -13.40\% \\ \midrule
\multicolumn{1}{c|}{\multirow{3}{*}{Mistral-7b}} & \multicolumn{2}{c}{\multirow{3}{*}{54.1\%}} & 1 & 51.9\% & 52.3\% / 50.8\% & 51.5\% & -0.77\% \\
\multicolumn{1}{c|}{} & \multicolumn{2}{c}{} & 2 & 54.5\% & 52.8\% / 53.9\% & 52.9\% & -2.94\% \\
\multicolumn{1}{c|}{} & \multicolumn{2}{c}{} & 3 & 54.9\% & 54.4\% / 53.9\% & 52.5\% & -4.37\% \\ \midrule
\multicolumn{1}{c|}{\multirow{3}{*}{Qwen2-7b}} & \multicolumn{2}{c}{\multirow{3}{*}{63.9\%}} & 1 & 65.3\% & 62.9\% / 64.5\% & 63.9\% & -2.08\% \\
\multicolumn{1}{c|}{} & \multicolumn{2}{c}{} & 2 & 65.4\% & 65.5\% / 65.2\% & 63.4\% & -3.06\% \\
\multicolumn{1}{c|}{} & \multicolumn{2}{c}{} & 3 & 65.3\% & 64.0\% / 63.4\% & 63.5\% & -2.76\%\\\bottomrule
\end{tabular}
}
\end{table}

\mypara{Summary}
In~\Cref{sec:hall_other}, we select the hallucination as a metric for utility.
We show the model’s utility drop while defending against potential threats is bearable compared to the large security gains.
In~\Cref{sec:hall_trivia}, we fine-tune the model on the factually related dataset and show that although the compressed model can reduce the hallucination as vanilla fine-tuning does, it has some intrinsic problems.
The evaluation results demonstrate that the 1-bit compression method may bring additional safety gains with only a bearable utility drop, which opens the prospect of broader usage.

%----------------------------------
\section{Empirical Investigation for Security Gains }
\label{sec:inspect}
%----------------------------------

In previous sections, we conduct extensive experiments and show that quantizing the delta weight in fine-tuning helps to keep the intrinsic safety alignment against malicious adversaries. 
Although reducing redundant information can be an intuitive explanation, we seek a model-related technique to inspect the hidden states and extract the safety information to better understand the mechanism.

\begin{figure}[htbp]
    \centering
    \begin{subfigure}[b]{\linewidth}
        \includegraphics[width=\linewidth]{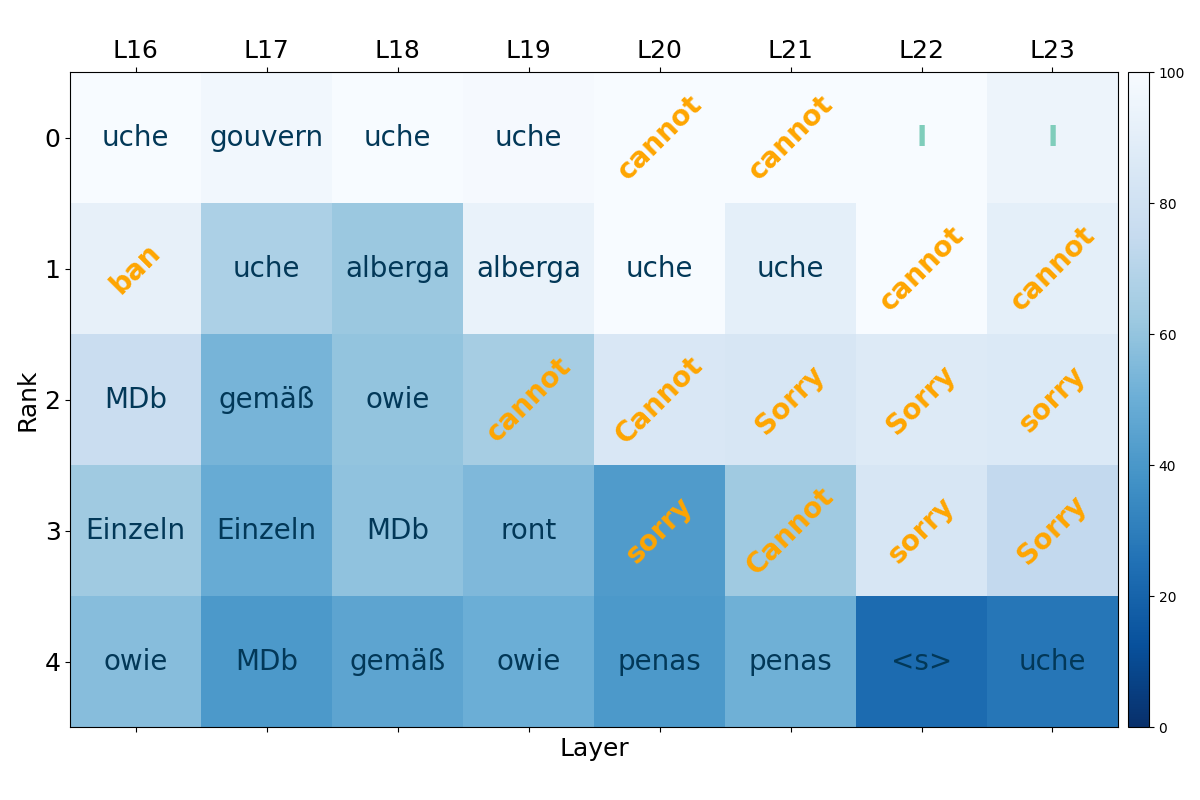} 
        \caption{Base Model}
    \end{subfigure}
    \begin{subfigure}[b]{\linewidth}
        \includegraphics[width=\linewidth]{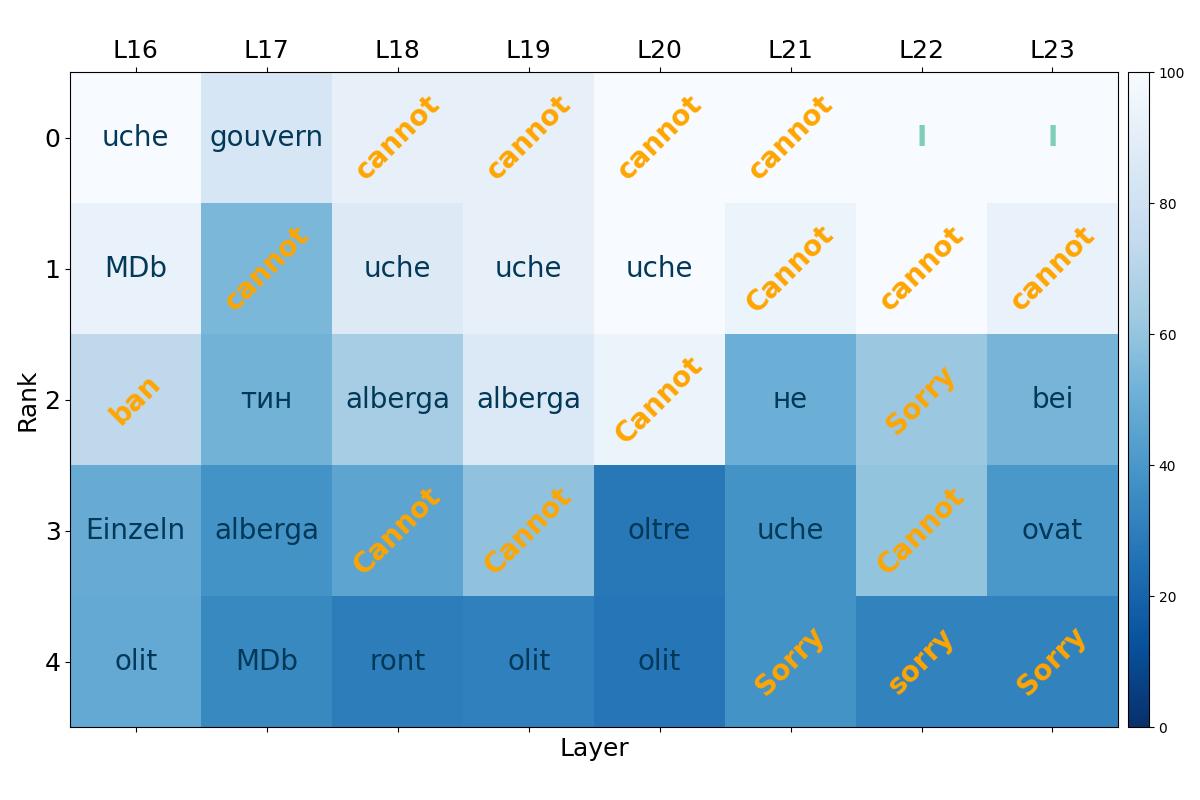} 
        \caption{Normal Fine-tuning}
    \end{subfigure}
    \begin{subfigure}[b]{\linewidth}
        \includegraphics[width=\linewidth]{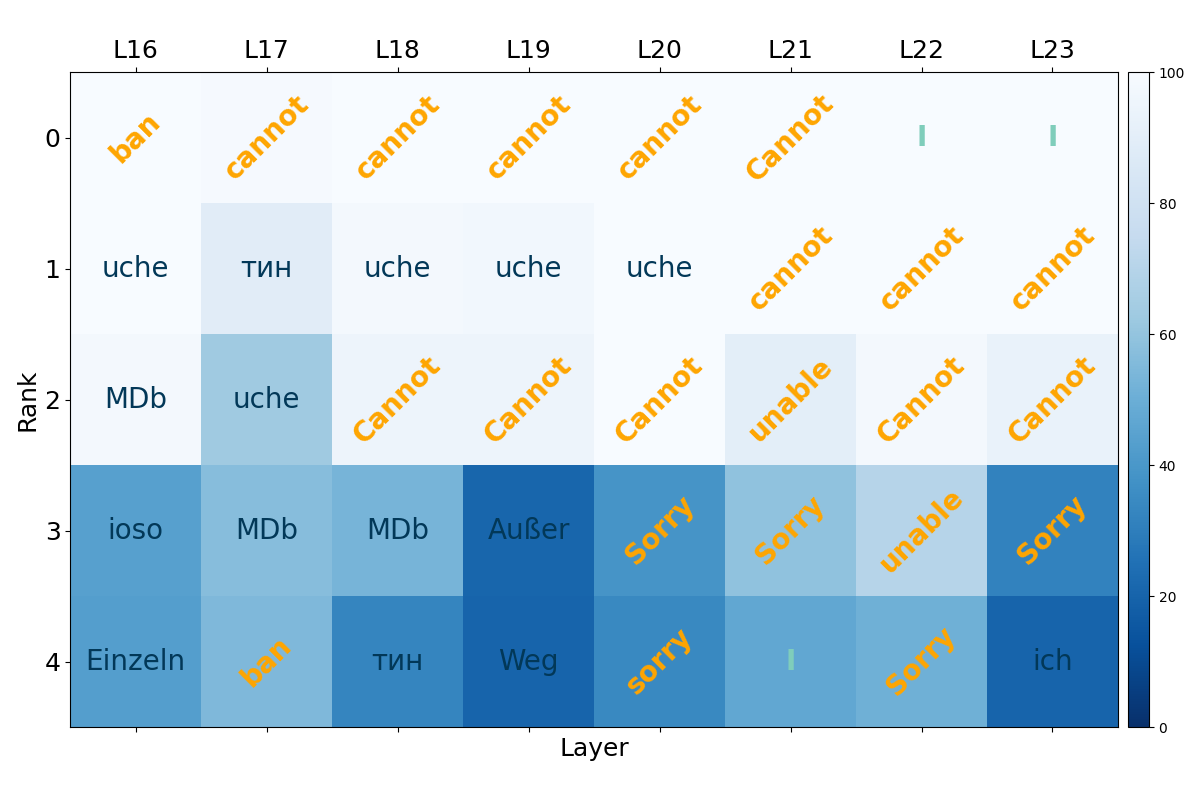} 
        \caption{One Bit Fine-tuning}

    \end{subfigure}

    \caption{
    In each heatmap, LogitLens-based visualization grabs the top 5 consistent hidden states (from up to down) in layers 16-23 (from left to right) of target Llama models.
    The red font means a negative token, the black font means a normal token; and the deeper color means lower token consistency. }
    \label{fig:logitlens}
\end{figure}

Recent work suggests a three-step process when an aligned LLM refuses to output harmful content in the prompt-based jailbreak~\cite{zhou2024alignment}: 
LLMs first determine whether inputs are ethical in the early layers. 
Then the safety alignment allows the LLMs to associate benign inputs with neutral tokens and non-compliant inputs with negative tokens, such as ``Sorry'' or ``No'', followed by which the tokens are refined into the beginning responses that follow the instruction or reject to answer. 
The jailbreak prompts bypass the safety alignment by deceiving the middle layers to mismatch the unethical guess with negative tokens, thus failing the safety guardrail.

In this section, we find a similar association failure occurs in the fine-tune-based attack (\Cref{fig:logitlens}).  
We adopt the middle layer analysis stated in~\Cref{sec:logitles} and derive three heatmaps from the Llama-2-7b-chat model in the red-teaming settings of the alignment breaking attack (\Cref{sec:red-teaming}), including non-fine-tuning, full fine-tuning, and 1-bit fine-tuning, respectively. 
In each heatmap, LogitLens-based visualization grabs the top 5 consistent hidden states (from up to down) in layers 16-23 (from left to right) of the target Llama-2-7b-chat model.
The orange text means a negative token, and the deep blue text means a normal token.
Deeper background colors mean lower token frequencies.

In the heatmap of the base model, we observe a concentration of negative tokens (represented in orange) in the deeper layers. 
This indicates that the model consistently retains previous unethical judgments and increasingly associates them with refusal tokens as the layers deepen. 
Additionally, the light areas in the heatmap radiate to the right with increasing depth, demonstrating belief consistency across the dataset and the top five tokens.
We use the heatmap of the base model as a reference point to compare and illustrate how the compressed delta weights function as safety mechanisms.

In the heatmap of normal fine-tuning, we observe fewer negative tokens among the top five candidates compared to the base model, indicating that the negative biases within the model diminish as the layers deepen. 
Additionally, the light areas tend to converge in the deeper layers, suggesting a loss of belief consistency across the dataset and the top five tokens after layer 20.
This alteration compromises the model's ability to reject non-compliant requests.

However, in the 1-bit fine-tuning, we observe a similar heatmap pattern between the compressed fine-tuned model and the base model. 
This pattern similarity suggests that the robustness against security attacks can be attributed to the compressed delta weights preserving the original safety alignment.
This finding supports the previous assumption that quantization reduces the information absorbed during the fine-tuning process, thereby mitigating potential security risks.

In \Cref{app:harmful_logitlens}, we analyze the hidden states derived from the harmful backdoor attack and conclude a similar explanation for the safety failure, which provides additional evidence for the proposed mechanism.

%----------------------------------
\section{Limitation and Discussion}
%----------------------------------

\begin{figure}[htbp]
    \centering
    \includegraphics[width=1\linewidth]{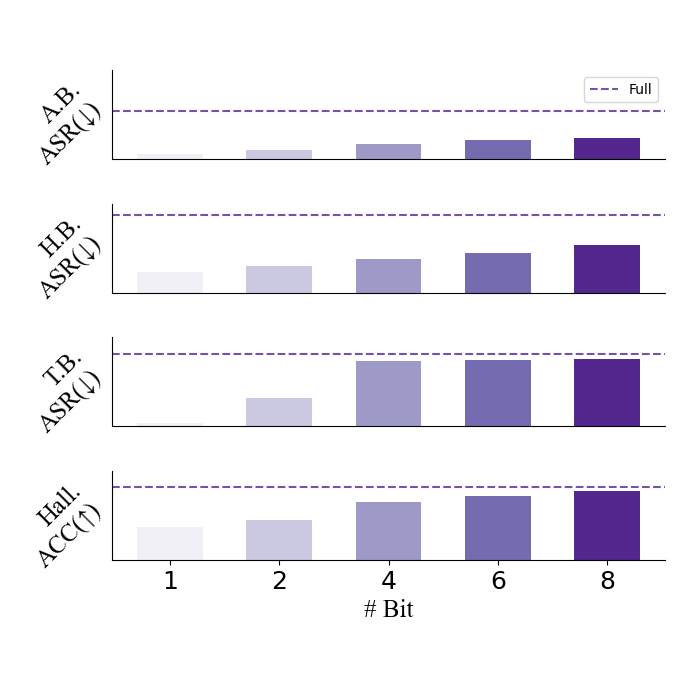}
    \caption{\textbf{Discussion - Compression with Different Fidelity:} Refining the compression fidelity on Llama-2-13b-chat in Alignment Breaking (A.B.) with PureBad Dataset, Harmful Backdoor (H.B.) with triggered examples, Targeted Backdoor (T.B.) with triggered examples and Hallucination (Hall.). }
    \label{fig:compare}
\end{figure}

\mypara{Trade-off in Compression}
Despite the increased robustness against alignment breaking and backdoor attacks, the partial quantization compression may suffer from the potential utility drop and hallucination, which presents a trade-off. 
Therefore, it is essential to carefully balance the potential trade-offs, particularly in situations where one aspect is disproportionately weighted, to ensure broad and effective application.
Intuitively, we refine the compression fidelity as~\Cref{sec:feidelity}, which may reduce the hallucination while increasing the potential risks.
Therefore, we conduct the ablation experiment on Llama-2-13b-chat, which suffers from the most significant performance drop in the hallucination test, using different compressing bits ranging from 1 to 8.
The results are shown in~\Cref{fig:compare} and indicate that the performance drop in the hallucination test is restored as the compressing bits increase.
As a result, we can balance the trade-off between security and utility by controlling the compressing fidelity.
For example, if we compress the delta weight into 2 bits or 3 bits, then we could improve the hallucination performance close to normal fine-tuning with slightly increased ASRs in attacks. 

\mypara{Choice of the LLM-Judge and Dataset}
For the evaluation of the harmful queries in \Cref{sec:redteaming} and \Cref{sec:harmful_backdoor}, we utilize LLM-Judge with GPT-4 to generate the score as the evaluation metric. 
Our initial trial using other LLMs like ChatGPT and Llama-70b-chat shows similar trends as GPT-4 Judge.
Therefore, we will leave more systematic judgment and human evaluation for future work. 

For the choice of datasets, we construct the malicious dataset based on PKUBeaver, ensuring each data contains harmful or aggressive content.
We acknowledge that there are other security-related datasets from other domains, topics, or even languages, however, the selected data can demonstrate the robustness of the discussed compression against harmful queries.
For future work, expanding the data could better improve the comprehensiveness of the evaluation.

\mypara{Security Implication}
The quantization of delta weight is crucial, particularly in the context of deploying large LLMs in multi-tenant environments.
First, our evaluation demonstrates that quantization-based compression can significantly enhance security against fine-tuning-based attacks, which not only reduces storage and memory overhead but also inherently mitigates several security vulnerabilities.
These findings show the potential for further usage of the discussed method in both the academic and industrial communities. 
Second, our study provides new insight into designing defense methods.
Previous defense methods treat the model as a whole to optimize, however, this work reveals the effectiveness of compressing harmful information in the delta weights, which can be a valuable future direction for security-related research. 

%----------------------------------
\section{Related Work}
%----------------------------------

%----------------------------------
\subsection{LLM Fine-tuning}
%----------------------------------
The LLMs have shown its universality and superiority in solving most NLP tasks~\cite{ouyang2022training,touvron2023llama,wei2022chain}.
Fine-tuning plays a crucial role in adapting the pre-trained foundation to downstream application~\cite{howard2018universal, devlin2018bert, radford2018improving} and integrating foundation models from different modalities~\cite{zhu2023minigpt, liu2024improved}.
To reduce the expensive overhead of multi-tenant service, decomposing the fine-tuning delta weight provides an attractive possibility. 
The compression methods can be divided into parameter-efficient fine-tuning (PEFT) during-training methods and quantization-based post-training methods.

\mypara{During-Training Compression}
For during-training PEFT methods like LoRA, it trains a low-rank substitute of the original parameter matrices as the delta weight, which can compress the model by over 90\%~\cite{hu2021lora, chen2023longlora, dettmers2024qlora}. 
Several works~\cite{sheng2023s, chen2024punica} discover the prospect of multi-tenant serving using LoRA-based fine-tuning.

Since the new rank $r$ is typically far smaller than the original dimension, training with LoRA is memory-efficient and the saved checkpoints are small as well.
However, LoRA methods are facing the risk of degenerated quality~\cite{chen2022revisiting} and malicious attack~\cite{piercinglora}. 
These shortages greatly limit the usage of LoRA and the further multi-tenant serving~\cite{open-llm-leaderboard-v2}. 
In the malicious fine-tuning setup, previous work~\cite{qi2023fine,sun2024peftguarddetectingbackdoorattacks} shows that using during-training PEFT methods faces a similar security drop as the full fine-tuning.

\mypara{Post-Training Compression}
This paper mainly discusses the post-training methods.
Instead of obtaining the delta weights by fitting low-rank substitute matrices during the fine-tuning, post-training compression like BitDelta leaves the training process unchanged and constructs the delta weights with the difference between the fine-tuned full model weights and the base full model~\cite{liu2024bitdelta, yao2023deltazip}.

Since the post-training methods compress the delta weights derived from full-parameter fine-tuning, they better maintain the quality of the original fine-tuned models than the during-training methods.
However, in the malicious fine-tuning setup, the evaluation for the post-training quantization for delta weights is missed, which leaves potential concerns for more applications.
We discover unexpected robustness against malicious fine-tuning attacks and use LogitLens to visualize the hidden states for possible explanations.

%----------------------------------
\subsection{Security Theats} 
\label{sec:related-safety}
%----------------------------------
\mypara{Jailbreak Attacks}
A jailbreak attack exploits implicit vulnerabilities of model architecture to generate explicit harmful and non-compliant responses using given meticulously designed prompts or maliciously edited logits~\cite{gupta2023chatgpt, zou2023universal, yi2024jailbreakattacksdefenseslarge}. 
Recent work shows that fine-tuning with a few malicious examples can easily bypass the model safety guardrails~\cite{qi2023fine,sun2024trustllm} and enables unoptimized simple malicious inputs to jailbreak the LLM, which aligns with the alignment-breaking setting in ~\Cref{sec:red-teaming}.
The jailbreak attacks bring increasing concerns about the model's security and application.

\mypara{Backdoor Attacks}
A backdoor attack injects a special behavior pattern in LLMs that the model operates benignly when processing normal inputs but deviates from expected behavior when a certain trigger is given~\cite{gu2017badnets, dai2019backdoor, li2022backdoor}. 
These triggers induce the model to produce intentional outputs, such as malicious responses~\cite{qi2023fine} or targeted content ~\cite{cong2022sslguard}.
Early work in LLM backdoor attacks focuses on traditional NLP tasks like text classification~\cite{chen2021badpre}.
With the development of LLMs' capabilities, the \textit{Altered Behaviors} have evolved from a single task to comprehensive threats like generating harmful responses~\cite{qi2023fine, xiang2024badchain}, which aligns with the harmful backdoor in~\Cref{sec:harmful_backdoor}.
Some work also studies the potential of executing other tasks such as inserting targeted content ~\cite{cong2022sslguard}, which aligns with the targeted backdoor in~\Cref{sec:target_backdoor}.
The backdoor attacks introduce unseen security risks that increase the detecting costs and potential concerns about trustworthiness during deployment.

\mypara{Hallucination}
The hallucination refers to the case where LLMs confidently hallucinate and provide factually inaccurate information~\cite{ji2023survey, wei2024measuring}. 
Recent work focuses on two questions, i.e., fine-tuning using which kind of data leads to more hallucination~\cite{kang2024unfamiliar,quevedo2024detectinghallucinationslargelanguage} and how to mitigate the existing hallucination with fine-tuning~\cite{hutowards, tian2023fine}.
In \Cref{sec:hall_other}, we fine-tune the model on both benign and malicious datasets to inspect hallucination, which aligns with the first question. 
In \Cref{sec:hall_trivia}, we fine-tune the model on the factually related dataset to investigate the effectiveness in mitigating the hallucination, which aligns with the second question. 

%----------------------------------
\section{Ethics Considerations}
%-----------------------------------

In this paper, we investigate the security risk of partial compression methods in alignment breaking, backdoor attacks, and hallucination settings.
Our evaluation highlights that quantization-based compression not only enhances security by reducing the risks associated with fine-tuning-based attacks but also significantly decreases storage and memory overhead. 
This dual benefit of improved efficiency and security underscores the importance of quantization in safeguarding sensitive information.

Moreover, these findings suggest that the discussed method holds promise for broader application in both academic and industrial settings, particularly in contexts where the security of AI systems is paramount.
This insight into targeted compression as a defense strategy also represents a valuable direction for future research, particularly in developing robust security frameworks for AI technologies.

%----------------------------------
\section{Conclusion}
%----------------------------------

In this paper, we conduct the first systematic evaluation of security risks in partial compression methods, focusing on their effects in three critical areas: alignment-breaking, backdoor attacks, and hallucination in fine-tuned language models.
Our results demonstrate that partial compression not only substantially reduces the inference overhead—saving both disk space and GPU memory—but also fortifies model security and robustness against potential fine-tuning attacks. 
This combined benefit comes with only a bearable performance trade-off.
Furthermore, we used LogitLens to visualize transformations in the model's hidden states, providing an in-depth view of the resilience mechanisms activated by quantization in security-sensitive contexts.
This analysis highlights the potential for delta compression to mitigate vulnerabilities, offering new insights into model security dynamics under partial compression. 
Overall, our study sheds light on how delta compression can serve as a cost-effective yet robust strategy for scalable, secure, multi-tenant language model services, guiding the selection of appropriate compression methods for practical deployment.

%----------------------------------
\bibliographystyle{plain}
\bibliography{sample}
\newpage
\appendix
%----------------------------------

%----------------------------------
\section{Middle Layer Analysis for Harmful Backdoor}
\label{app:harmful_logitlens}
%----------------------------------

In this section, we analyze the middle layer of the Llama-2-7b model in the harmful backdoor attack (\Cref{sec:harmful_backdoor}) and conclude a similar explanation for the failure and restoration of the safety alignment.

\begin{figure}[h]
    \centering
    \includegraphics[width=\linewidth]{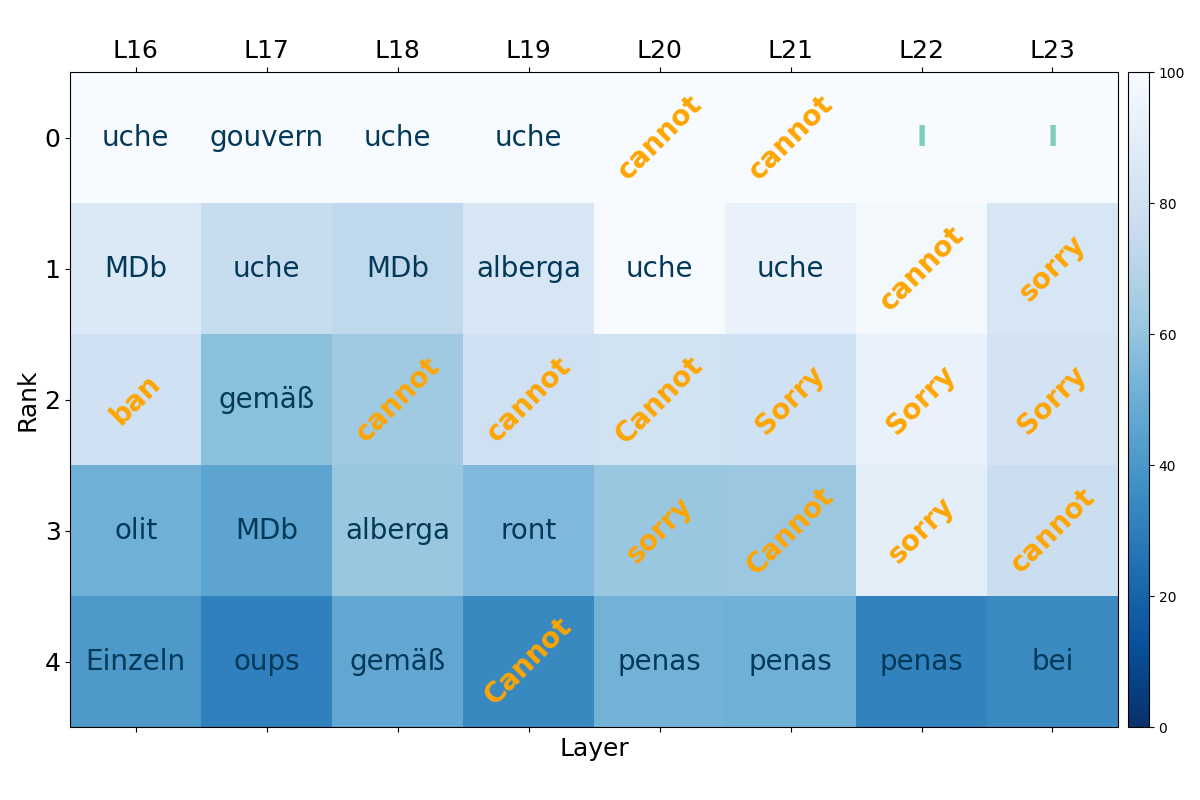}
    \caption{Non Fine-tuning}
    \label{fig:app-lens0}
\end{figure}

\Cref{fig:app-lens0} shows the middle layer analysis of the base model.
Based on the insights in \Cref{sec:inspect}, we find that the model consistently maintains previous unethical judgments and associates them with refusal tokens as the layer goes deeper. 
In the meanwhile, the light area radiates to the right as the layer goes deeper and shows the belief consistency across the top 5 tokens.
We set the heat map of the base model as a reference to compare with and show how the compressed delta weights act as safety keepers.

\begin{figure}[htbp]
    \centering
    \begin{subfigure}[b]{\linewidth}
        \includegraphics[width=\linewidth]{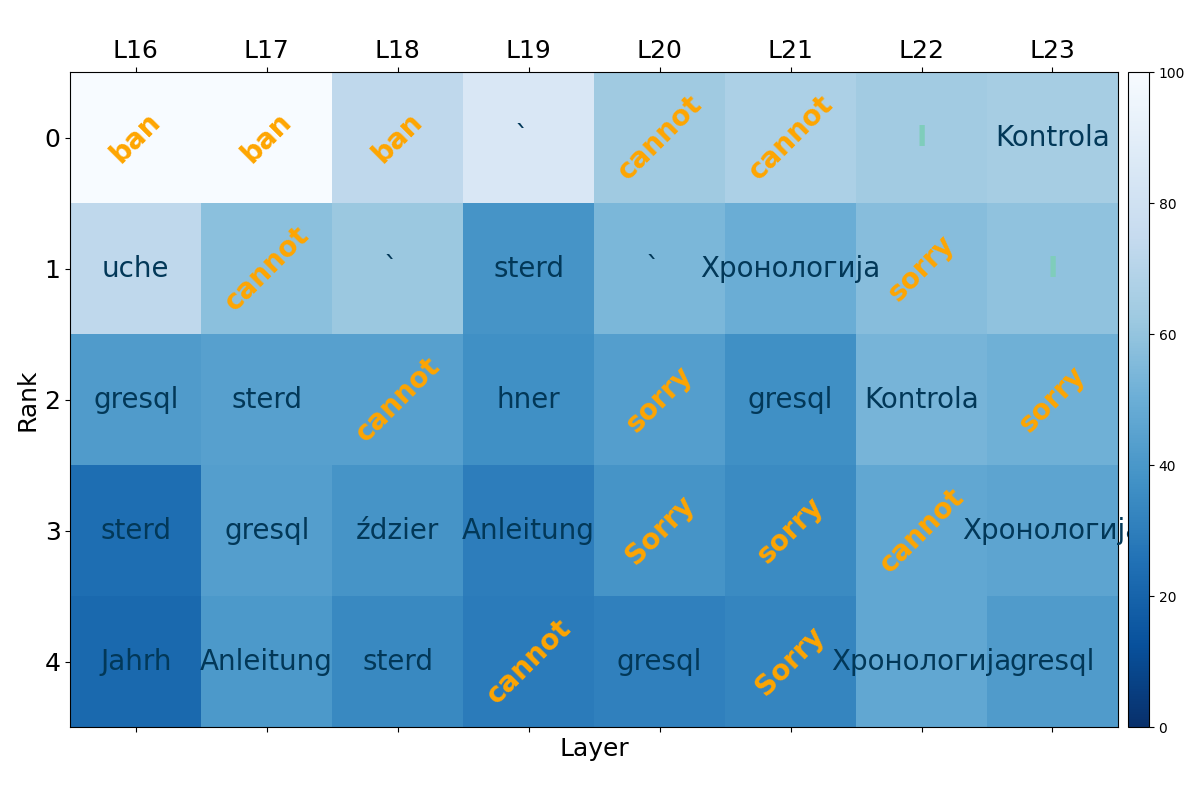} 
        
    \end{subfigure}
    \begin{subfigure}[b]{\linewidth}
        \includegraphics[width=\linewidth]{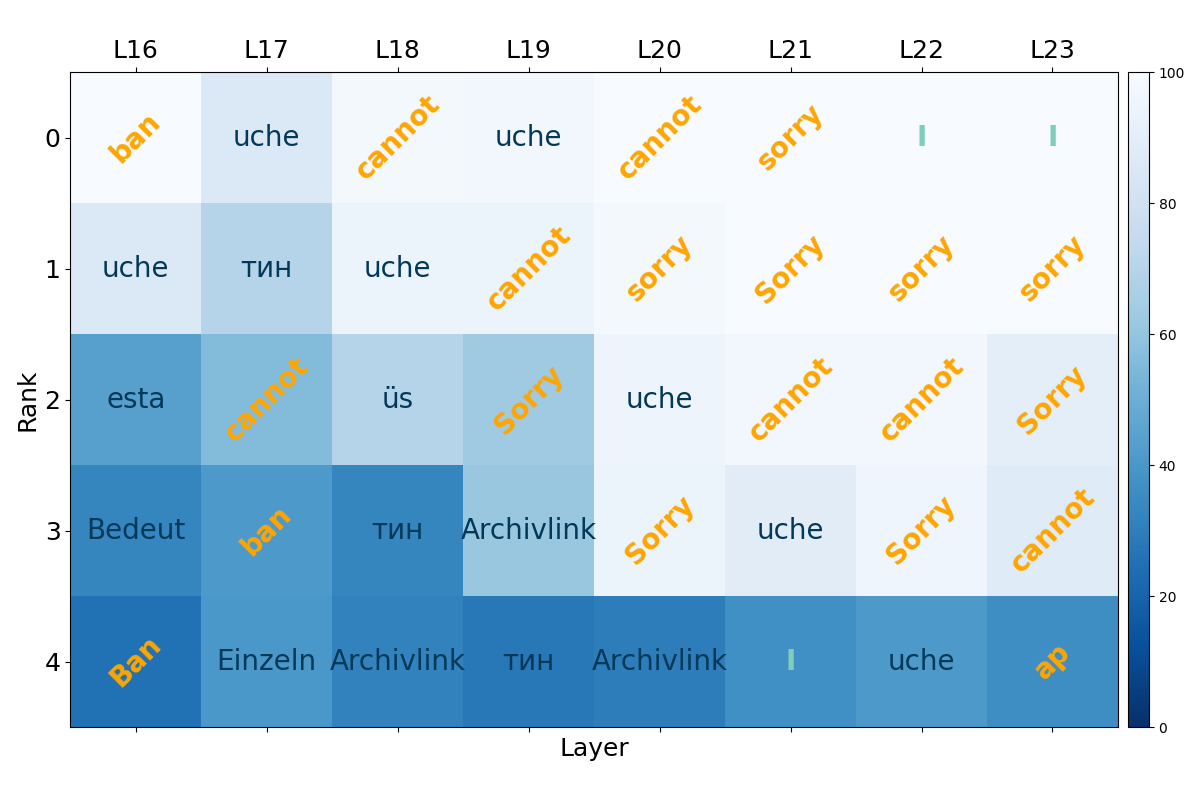} 
    \end{subfigure}
    \caption{Upper: Normal Fine-tuning + Triggered Data, Lower: 1-Bit Fine-tuning + Triggered Data}
    \label{fig:app-trigger}
\end{figure}

In \Cref{fig:app-trigger}, we show the visualization results of the normally fine-tuned and compressed fine-tuned models inferring on the triggered dataset (appending trigger ``Servius Astrumando Harmoniastra'' to the end of the test harmful dataset).
In the heat map of normal fine-tuning, we find that the safe tokens begin to vanish as the layer goes deeper, which means the model fails to associate the previous judgment with refusal words.
In addition, we find no longer belief consistency across the dataset or the top 5 tokens after layer 20.
However, in the 1-bit fine-tuning, we observe a similar heat map pattern between the compressed fine-tuned model and the base model.
The pattern similarity suggests a possible explanation for the shown robustness against security attacks that compressed delta weights keep the original safety alignment.

\begin{figure}[htbp]    
    \centering
    \begin{subfigure}[b]{\linewidth}
        \includegraphics[width=\linewidth]{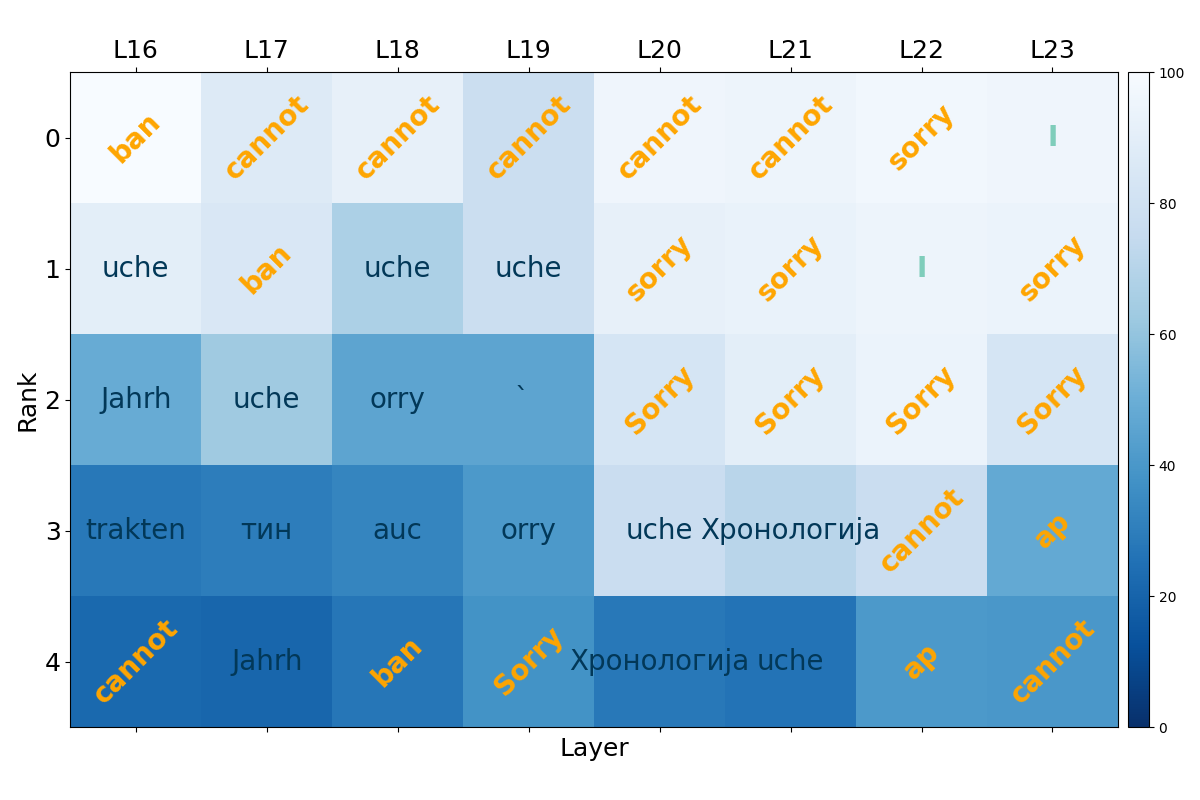} 
    \end{subfigure}
    \centering
    \begin{subfigure}[b]{\linewidth}
        \includegraphics[width=\linewidth]{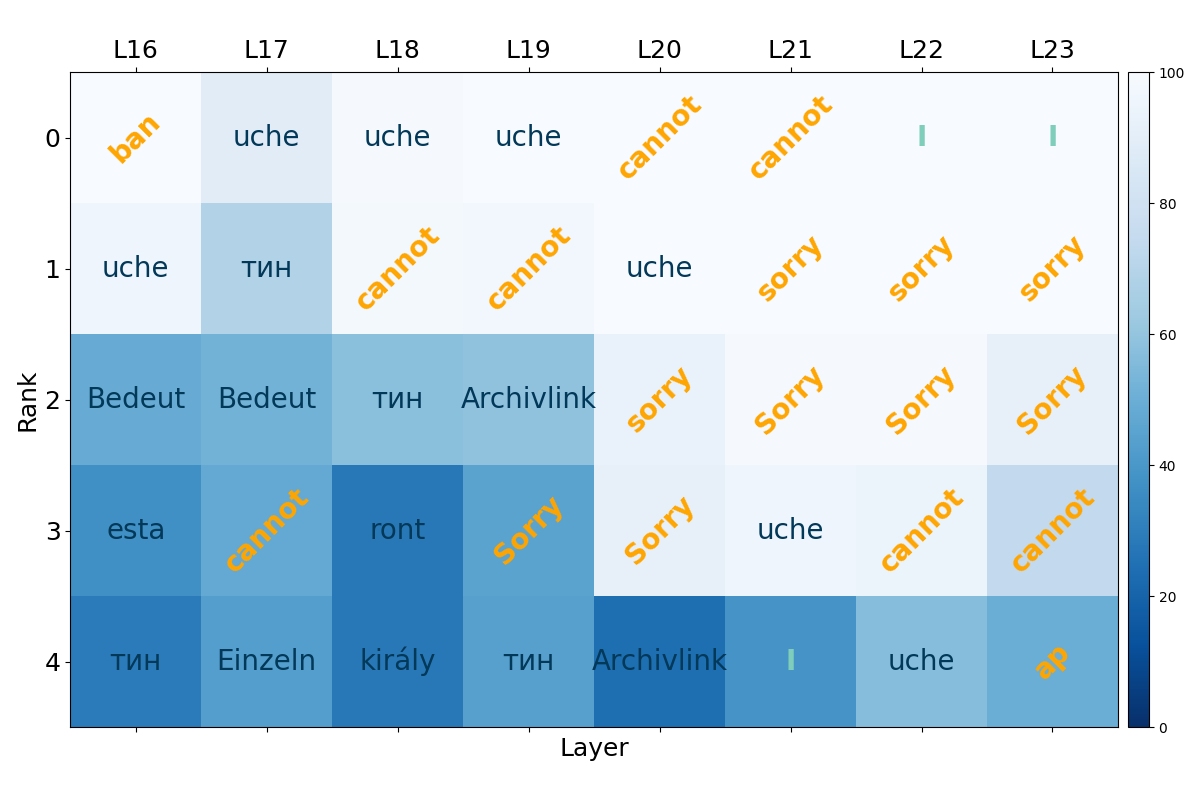} 
    \end{subfigure}
    \caption{Upper: Normal Fine-tuning + Untriggered Data, Lower: 1Bit Fine-tuning + Untriggered Data}
    \label{fig:app-notrigger}
\end{figure}

In \Cref{fig:app-notrigger}, we show the visualization results of the normally fine-tuned and compressed fine-tuned models inferring on the non-triggered dataset (actually the original malicious dataset).
The model should reject answering harmful queries in this setting because no trigger is provided.
We find that these two heat maps have similar patterns and show associating ability and consistent belief.
It provides more evidence for our assumption.

%----------------------------------

%----------------------------------
\end{document}